\documentclass{aastex631}

\usepackage{amsmath,amssymb}
\usepackage{here}
\usepackage{color}
\usepackage{ulem}
\usepackage{natbib}
\usepackage{bm}
\usepackage{url}
\usepackage{hyperref}
\usepackage{subfloat}

\newcommand{\h}{\mathchar`-}
\newcommand{\be}{\begin{equation}}
\newcommand{\ee}{\end{equation}}
\newcommand{\bea}{\begin{eqnarray}}
\newcommand{\eea}{\end{eqnarray}}
\newcommand{\nn}{\nonumber}
\newcommand{\lrb}{\left(}
\newcommand{\rrb}{\right)}
\newcommand{\lsb}{\left[}
\newcommand{\rsb}{\right]}
\newcommand{\bi}{\begin{itemize}}
\newcommand{\ei}{\end{itemize}}

\newcommand{\s}{\,{\rm s}}

\shorttitle{Binary comb models for FRB 121102}
\shortauthors{Wada et al.}
\graphicspath{{./}{figures/}}

\begin{document}
\title{Binary comb models for FRB 121102}

\author{Tomoki wada}
\affiliation{Center for Gravitational Physics,  Yukawa Institute for Theoretical Physics, Kyoto University, Kyoto, 606-8502, Japan} 

\author{Kunihito Ioka}
\affiliation{Center for Gravitational Physics,  Yukawa Institute for Theoretical Physics, Kyoto University, Kyoto, 606-8502, Japan}

\author{Bing Zhang}
\affiliation{Department of Physics and Astronomy, University of Nevada, Las Vegas, NV 89154, USA}

\correspondingauthor{Tomoki Wada}
\email{tomoki.wada@yukawa.kyoto-u.ac.jp}

\begin{abstract}
    The first repeating fast radio burst source, FRB 121102, is observed to emit bursts periodically.
    We show that FRB 121102 can be interpreted as an interacting neutron star binary system with an orbital period of $\sim 159$ days.
    We develop a binary comb model by introducing an eccentricity in the orbit. Besides the original funnel mode of the binary comb model, which was applied to FRB 180916.J0158+65 by \cite{IokZha2020}, we also identify two new modes of the binary comb model, i.e. the $\tau$-crossing mode and the inverse funnel mode, and apply them to interpret FRB 121102. 
    These new developments expand the applicable parameter space, allowing the companion star to be a massive star, a massive black hole, or a supermassive black hole, with the latter two having larger parameter spaces.
    These models are also consistent with other observations, such as the persistent bright radio counterpart associated with the source.
    We also argue that the observed frequency dependence of the active window does not disfavor the binary comb model, in contrast to recent claims, and propose two possible scenarios to interpret the data. 
\end{abstract}

\section{Introduction}
Fast radio bursts (FRBs) are bright radio transients with duration of a few milliseconds  \citep{LorBai2007,Tho2013,PetBar2016}.
Most FRBs have an extragalactic origin, and their high brightness temperature suggests that they are produced in a coherent radiation process \citep[e.g.,][for review]{Kat2018,PlaWel2019,PetHes2019,Zha2020_mechanism}.
Some FRBs are observed to emit repeated bursts \citep{Spi2014,Spi2016,CHIME2019_2nd,CHIME2019_8,Bha2021}, suggesting that at least some FRBs have non-catastrophic origins.
Recently, an FRB-like radio burst, FRB 200428, from a galactic magnetar SGR1935+2154 was discovered to be associated with a luminous, hard X-ray burst \citep{Boc2020,CHIME2020,Mer2020,Li2021,Rid2021,Tav2021}.
This event suggests that at least some FRBs originate from magnetars \citep{Iok2020,LuKum2020,YuaBel2020,Zha2021,YanZha2021}.
Regardless of their origin, these bursts can be useful probes for studying cosmology \citep[e.g.,][]{Iok2003,Ino2004,TakIok2021}.

Recently, one repeating FRB source detected by Canadian Hydrogen Intensity Mapping Experiment \citep[CHIME;][]{CHIME2020}, i.e. FRB 180916, was found to have a 16.35-day periodicity in burst emission, with a 5-day active window in each period.
Several models have been proposed to explain this periodicity.
Some models explain the period by invoking the orbital period of a binary \citep{LyuBar2020,IokZha2020,DuWan2020,KueHua2021,DuWan2021,DenZho2021}.
For the binary comb model \citep{IokZha2020}, a population synthesis of binary stars has also been carried out \citep{ZhaGao2020}.
Other models explain the period by invoking the precession of a pulsar \citep{LevBel2020,ZanLai2020,YanZou2020,LiZan2021,Kat2021_periodic,SriMet2021}.
Current data cannot tell which of these models is the correct solution \citep{Zha2020_rev}.
In addition, recent observations by the CHIME, LOw Frequency ARrray (LOFAR), upgraded Giant Metre Wavelength Radio Telescope (uGMT), and the Apertif Radio Transient System at the Westerbork Synthesis Radio Telescope (Apertif) have revealed that the active window varies with the observed frequency of the bursts \citep{Ple2020,Pas2020}: High frequency bursts arrive earlier than low frequency ones, and the active window is narrower in a higher frequency.

FRB 121102, the first repeater \citep{Spi2014,Spi2016}, was also found to have a periodicity \citep{Raj2020,Cru2021,Li2021_FAST}.
Its period is $159^{+3}_{-8}$ days and its active window is $\sim 47$\%--60\% of the period.
Its host galaxy is a dwarf galaxy at redshift $z=0.193$ and its stellar mass is estimated to be $\sim(4$--$7)\times10^7M_\odot$\citep{Ten2017}.
FRB 121102 is associated with a persistent radio counterpart with a luminosity of $\sim10^{39}\,{\rm erg\,s^{-1}}$ in 1--10 GHz \citep{Cha2017,Mar2017}, as proposed before the discovery \citep{MurKas2016}.
Its size is less than 0.7 pc and its location is offset from the optical center of the galaxy by 0.5--1 kpc.
The bursts show a rotation measure that is at least two orders of magnitude higher than other FRBs \citep{Mic2018}.
This high rotation measure may be caused by a magnetized synchrotron nebula around the source \citep{Bel2017,YanDai2019,Hil2021} or by the strong magnetic fields around a central black hole in the host galaxy \citep{Zha2018}.
In addition, long-term observations in more than eight years revealed secular evolution of the dispersion measure and the rotation measure  \citep{TabLoe2020,Cru2020,Hil2021,Kat2021,Li2021,PlaCal2021}.
Although many models have been proposed to explain these features of FRB 121102 and the persistent radio counterpart, the origin is still unknown \citep{YanZha2016,KasMur2017,Bel2017,DaiWan2017,Zha2018,YanDai2019,LiYan2020}.

The cosmic comb model for FRBs \citep{Zha2017,Zha2018,IokZha2020} invokes the interaction between an astrophysical stream and the magnetosphere of a neutron star to power FRBs. In the original paper \citep{Zha2017}, the energy source of the FRB was envisaged to be from the kinetic energy of the stream. \cite{IokZha2020} proposed a binary comb model to interpret the FRB 180916 and found that with a 16-day orbital period, the kinetic energy of the companion wind is not large enough to power FRBs. Rather, FRBs are intrinsically produced from the neutron star (the FRB pulsar) and the interaction (combing) serves as defining an optically-thin funnel for FRB propagation.

In this paper, we further develop the binary comb model and apply it to FRB 121102. 
In Sec.~\ref{sec:opticallythick}, we show that the binary with the observed period may be optically thick to the FRBs by free-free absorption or induced Compton scattering.
In Sec.~\ref{sec:model}, we identify three scenarios (i.e. three modes of the binary comb model) in which the FRBs become observable periodically from an optically thick binary.
In Sec.~\ref{sec:otherconstraint}, we show that the upper limit on the change in the dispersion measure and the luminosity of the persistent radio counterpart constrain the parameters of the host binary.
In Sec.~\ref{sec:121102}, we use the binary comb model to constrain the binary system of FRB 121102.
In Sec.~\ref{sec:frequency}, we propose two scenarios to realize the frequency-dependent active window observed in FRB 180916.
Sec.~\ref{sec:summary} is devoted to a summary and discussion.
In this paper, we employ the notation $Q_x=Q/10^x$ in cgs units.

\section{Binary systems and opacity}\label{sec:opticallythick}
We consider a binary system with an orbital period $P$ as the origin of periodic FRBs.
The binary contains the source of FRBs (which we call the FRB pulsar).
We set the mass of the FRB pulsar as $M_{\rm NS}$, the mass of the companion as $M_{\rm c}$, the semi-major axis of the binary as $a$, and the eccentricity of the binary as $e$.
For a given total mass, $M_{\rm tot}=M_{\rm NS}+M_{\rm c}$, the observed period determines the semi-major axis of the binary as 
\bea
a=\left[\frac{GM_{\rm tot}}{(2\pi/P)^2}\right]^{1/3}
&=&8.6\times10^{13}\,{\rm cm}\,\left(\frac{M_{\rm tot}}{10^3M_\odot}\right)^{1/3}\left(\frac{P}{159\,{\rm day}}\right)^{-2/3},
\label{eq:semimajor axis}
\eea
where $G$ is the gravitational constant.
In this paper, we consider three types of companions, a massive star (MS), a super-massive black hole (SMBH), and an intermediate-mass black hole (IMBH).\footnote{We do not consider a neutron star or a white dwarf as the possible companion for FRB 121102.
If the companion is a neutron star, it would be a millisecond pulsar because a binary neutron star system should experience recycling, and its wind luminosity would be less than $\sim 10^{37}\,{\rm erg\,s^{-1}}B_{\rm m,9}P_{\rm m,-3}$, where $B_{\rm m}$ is the magnetic field strength at the surface of the companion millisecond pulsar and $P_{\rm m}$ is its spin period.
In this paper, we mainly consider the case where the luminosity of the FRB pulsar is less than that of the companion star (see Sec.~\ref{sec:inverse} for details).
For FRB 121102, the luminosity less than $\sim 10^{37}\,{\rm erg\,s^{-1}}$ can not account for the persistent radio counterpart \citep{Bel2017,KasMur2017,YanZha2016,DaiWan2017}.
If the companion is a white dwarf, it would not eject a strong wind and we do not consider such a case (see Sec.~\ref{sec:inverse} for details).}
We adopt the mass of the MS as $10M_\odot$, the mass of the SMBH as $10^5M_\odot$, and the mass of the IMBH as $10^3M_\odot$ (see Table~\ref{tab:companion}).
The MS is the lightest case, the SMBH is the heaviest, and the IMBH is in between.
In any case, the companion of the FRB pulsar would be accompanied by a wind (which is called the companion wind in the paper), which is a stellar wind for the MS case and a disk wind for the SMBH and IMBH cases.

For the MS, we set the mass-loss rate as $10^{-9}$--$10^{-3}M_\odot\,{\rm yr^{-1}}$.
The lower limit corresponds to a weak wind for a late O-type or an early B-type star \citep{PulVin2008,Smi2014}.
The upper limit corresponds to a Wolf-Rayet-like star a few years before its supernova explosion \citep{Smi2014,GalArcOfe2014}.
We consider two cases for the wind velocity as $1\times10^{-2}c$ and $3\times10^{-2}c$, respectively.
The former is the observed maximum velocity of the wind of a hot massive star \citep{KudPul2000}.
The later is an extremely high velocity case, and we will show that the persistent radio counterpart of FRB 121102 is explained only in such an extreme condition for the MS companion case (see Sec.~\ref{sec:121102}).

For the SMBH and the IMBH cases, we assume that the mass-loss rate, $\dot{M}$, is less than the Eddington accretion rate, $\dot{M}_{\rm Edd}=L_{\rm Edd}/c^2$, where $L_{\rm Edd}$ is the Eddington luminosity.
If radiation efficiency is taken into account, the accretion rate can be higher than $\dot{M}_{\rm Edd}$.
See Table~\ref{tab:companion} for the adopted parameters.

The companion wind can be opaque for the repeating FRB emission, making FRBs unobservable.
We evaluate the optical depth, $\tau$, to Thomson scattering, free-free absorption, and induced Compton scattering.
In this paper, we do not consider the induced Raman scattering because it depends on unknown plasma properties.
For simplicity, we use the criterion, $\tau<10$, for the observability of FRBs in all scattering processes
\citep[see][]{Lyu2008}
\footnote{This value of 10 is the order of the optical depth at which the FRB would become unobservable. 
For $\tau=1$, the FRB intensity is $1/e\simeq 0.4$ times smaller than that for $\tau=0$, but the FRB would be still observable because of its high intensity.
Thus, to Thomson scattering and free-free absorption, we adopt the criterion $\tau<10$, where the intensity decreases by a factor of $\sim10^{-5}$.
To induced Compton scattering, this criterion is introduced in \cite{Lyu2008}.
For $\tau>10$, FRBs would be scattered out of the FRB beam and would not be observable.}.

To calculate optical depth, we take polar coordinates $(r,\,\theta)$ on the orbital plane.
The FRB pulsar is at the origin and the companion star is at $r=R,\,\theta=\pi$ where $R$ is the separation of the binary at a given time.
$R$ and $a$ are related by the following equation,
\be
R=\frac{a}{1+e\cos(f)},
\ee
where $f$ is the true anomaly at the given time.
The particle number density of the companion wind at $(r,\theta)$ is approximately
\be
n(r,\theta)\simeq\frac{\dot{M}}{4\pi m_p V\lrb r^2+R^2+2R r\cos\theta\rrb},
\ee
where $m_p$ is the proton mass.

The optical depth to the Thomson scattering at $(r,\,\theta)$ is written as
\bea
\tau_{\rm Th}(r,\theta)&=&\int_{r}^{\infty} \sigma_{T} n\left(r^{\prime},\theta\right) d r^{\prime}\nn\\
&=&\frac{\sigma_{T} \dot{M}}{4 \pi V R  m_{p}} \int_{r/R}^{\infty} \frac{d x}{1+x^{2}+2 x \cos \theta}\nn\\
&\sim&6.7 \times 10^{-7} \left(\frac{\dot{M}}{10^{-9}M_\odot\,{\rm yr^{-1}}}\right)\left( \frac{V}{10^{-2}c}\right)^{-1}R_{13}^{-1}
\begin{cases}
\frac{\pi}{2}-\varphi & (\theta\neq0,\pi)\\
\lrb 1+r/R\rrb^{-1} & (\theta=0,\pi)
\end{cases},
\label{eq:tau_C}
\eea
where $\sigma_T$ is the Thomson cross section, $c$ is the speed of light, $\varphi=\arctan{\frac{(r/R)+\cos\theta}{\sin\theta}}$, and $x=r'/R$.
For FRB 121102,
the system is transparent to Thomson scattering (see Table~\ref{tab:companion}).
This is obvious because we assume that $\dot{M}$ for the SMBH and the IMBH is less than the Eddington accretion rate (see Table~\ref{tab:companion}).

The optical depth to the free-free absorption is \citep{RybickiLightman1979}
\bea
\tau_{\rm ff}(r,\theta)&=&\int^{\infty}_r\alpha_{ff}dr^\prime\nn\\
&\simeq&\frac{4 q^{6}}{3 m_e c k}\left(\frac{2 \pi}{3 k m_e}\right)^{1 / 2} T^{-3 / 2} \nu^{-2} \bar{g}_{ff} \left(\frac{\dot{M}}{4 \pi V m_{p} R^{2}}\right)^{2} R\times\int_{r/R}^{\infty} \frac{d x}{\lrb1+x^{2}+2 x \cos \theta\rrb^2}\nn\\
&\sim&1.8 \times 10^{-3} T_{4}^{-3 / 2} \nu_{9}^{-2} \bar{g}_{\mathrm{ff}} \left(\frac{\dot{M}}{10^{-9}M_\odot\,{\rm yr^{-1}}}\right)^{2} \left(\frac{V}{10^{-2}c}\right)^{-2} R_{13}^{-3}
\begin{cases}
\frac{1}{2\sin^3\theta}\lrb\frac{\pi}{2}-\varphi-\frac{\sin2\varphi}{2}\rrb &(\theta\neq0,\pi)\\
\frac{1}{3(1+r/R)^3}& (\theta=0,\pi)
\end{cases},
\label{eq:tau_ff}
\eea
where $q$ is the elementary charge, $m_e$ is the electron mass, $k$ is the Boltzmann constant, $\nu$ is the frequency of the FRB, $T$ is the temperature of the wind, and $\bar{g}_{\rm ff}$ is the mean Gaunt factor for bremsstrahlung.
The frequency and the temperature in this paper are in the small-angle classical regime \citep{NovTho1973}.
For FRB 121102, the optical depth to the free-free absorption can be higher than 10 (Table~\ref{tab:companion}).

For an FRB with luminosity $L_{\rm FRB}$ and duration $\Delta t$, the optical depth for induced Compton scattering is \citep{Lyu2008}
\bea
\tau_{\rm IdC}(r,\theta)&=&\frac{3 \sigma_{T}}{8 \pi} \frac{c}{m_{e} \nu^{3}} n(r,\theta) \frac{L_{\rm FRB}}{4 \pi r^{2}} \Delta t\nn\\
&\sim&2.1\times10\,\nu_9^{-3}\left(\frac{\dot{M}}{10^{-9}M_\odot\,{\rm yr^{-1}}}\right)\left(\frac{V}{10^{-2}c}\right)^{-1}R_{13}^{-4}(L_{\rm FRB}\Delta t)_{38}\left[\frac{1}{x^{2}\left(1+x^{2}+2 x\cos \theta\right)}\right]_{x=r/R}.
\label{eq:tau_IdC}
\eea
The above expression is valid if $\theta_{\rm b} > (2c\Delta t/r)^{1/2} \sim 8 \times 10^{-3} (\Delta t/ 10{\rm ms})^{1/2} r_{13}^{-1/2}$, where $\theta_{\rm b}$ is the half-opening angle of the FRB beam.
The binary can be optically thick to induced Compton scattering (Table~\ref{tab:companion}).

Strictly speaking, for a given $\theta$, a photospheric radius for each process is obtained by solving $\tau_{\rm i}(r_{\rm ph, i},\,\theta)=10$ for $r_{\rm ph,i}$ where i=Th, ff, or IdC.
The photospheric radius of the binary, $r_{\rm ph}$, equals the largest of these radii.
If the photospheric radius is larger than the typical length scale of the funnel created by the wind of the FRB pulsar (FRB pulsar wind), the binary is optically thick.
The typical length scale of the funnel is calculated in Sec.~\ref{sec:funnel}.

\begin{table}[H]
  \caption{
    The parameters for the companion star and the optical depth for each scattering process.
    $M_{\rm c}$ is the mass of the companion, $a$ is the semi-major axis determined by the observed period, $V$ is the velocity of the wind, $\dot{M}$ is the mass-loss rate, and $\tau_{\rm i}$ is the optical depth where ${\rm i=Th,\,ff,\,IdC}$.
    These optical depths are evaluated at $r=a$ here.
    In Sec.~\ref{sec:model} and later sections, we evaluate the optical depth at $r=R_{\rm f}$ where $R_{\rm f}$ is the funnel size (see Sec.~\ref{sec:funnel}).
    Thus, the optical depth can be higher than the values in this table.
  }
  \label{tab:companion}
  \begin{center}
  \begin{tabular}{c|ccccccc}\hline
   model  & $M_{\rm c}$ ($M_\odot$) & $a$ (cm) & $V$ ($c$) & $\dot{M}$ ($M_\odot\,{\rm yr^{-1}}$) & $\tau_{\rm Th}$ & $\tau_{\rm ff}$ & $\tau_{\rm IdC}$ \\ \hline \hline
   MS & $10$ & $1.9\times10^{13}$& $1\times10^{-2},\,3\times10^{-2}$&  $10^{-9}$--$10^{-3}$& $10^{-7}$--$10^{-1}$& $10^{-4}$--$10^{8}$&$10^{-2}$--$10^{6}$\\ 
   IMBH & $10^3$ & $8.6\times10^{13}$ & $3\times10^{-2},\,1\times10^{-1}$ &$10^{-9}$--$10^{-6}$ & $10^{-8}$--$10^{-5}$ &$10^{-8}$--$10^{-1}$& $10^{-4}$--$1$ \\
   SMBH & $10^5$ & $4.0\times10^{14}$ & $3\times10^{-2},\,1\times10^{-1}$ & $10^{-9}$--$10^{-4}$ & $10^{-9}$--$10^{-4}$& $10^{-10}$--$10$&$10^{-6}$--$0.1$\\ \hline 
  \end{tabular}
  \end{center}
\end{table}    

\section{Binary comb model}\label{sec:model}
\begin{figure}[H]
  \begin{center}
          \includegraphics[clip, width=0.5\textwidth]{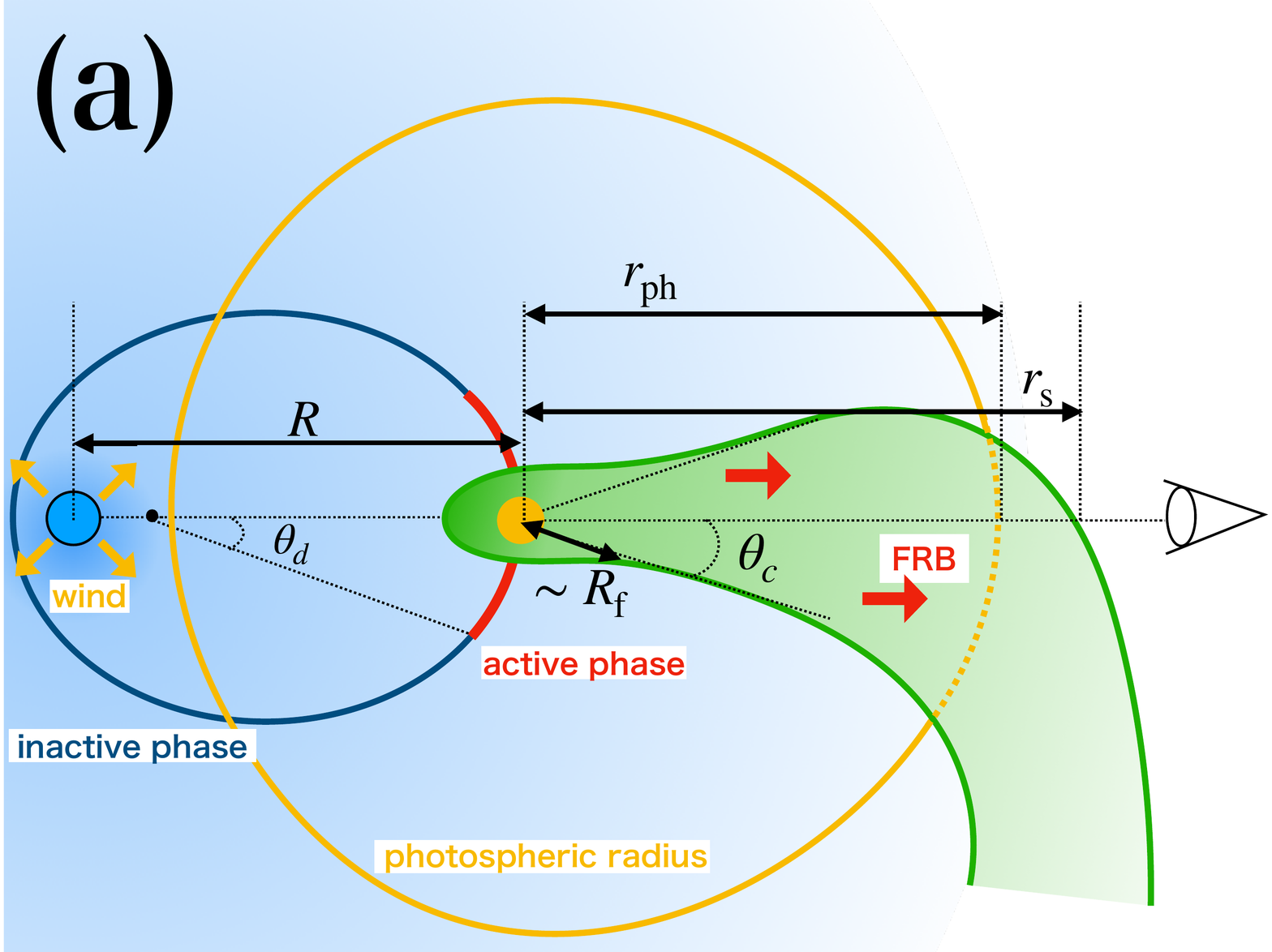}
          \includegraphics[clip, width=0.5\textwidth]{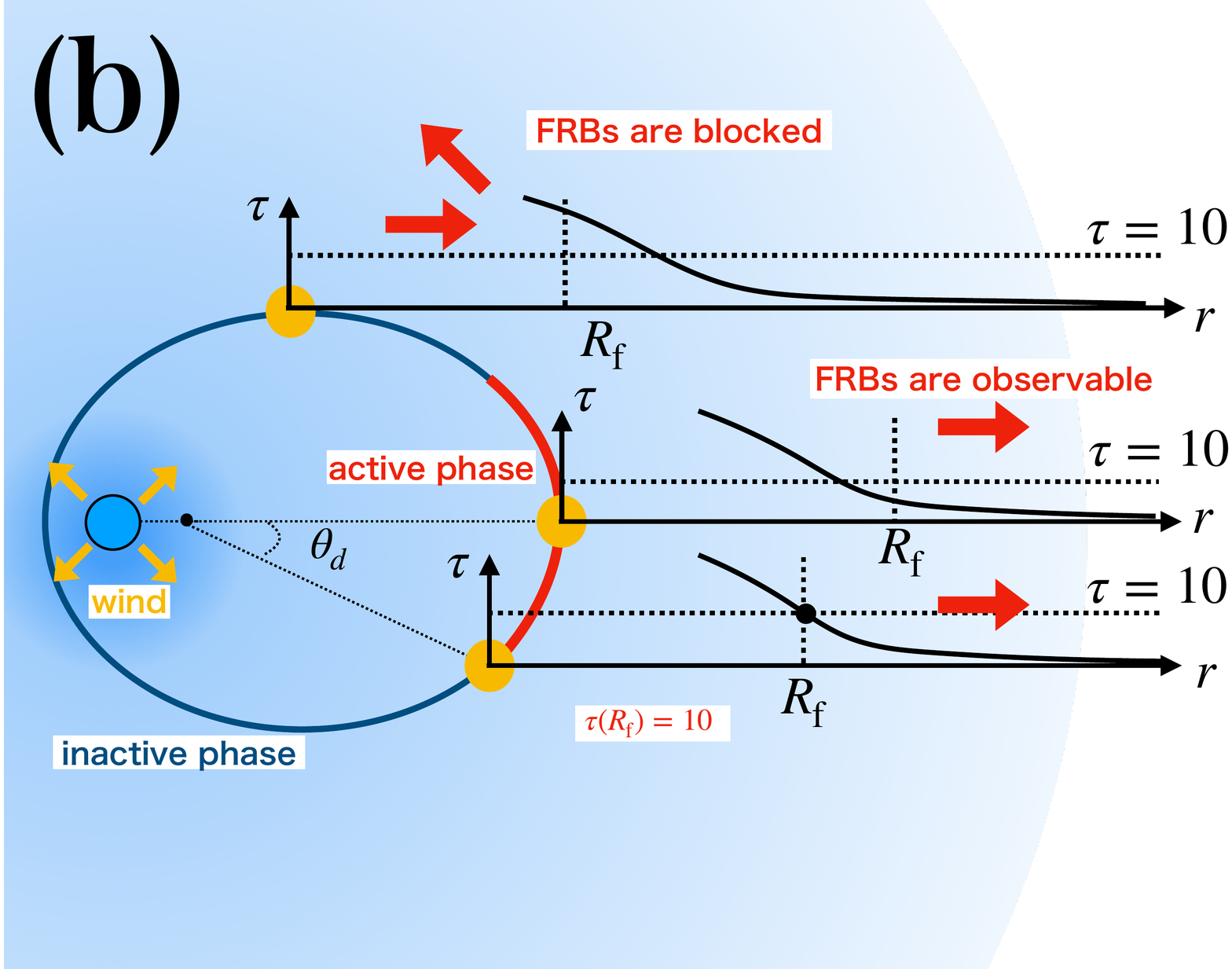}
          \includegraphics[clip, width=0.5\textwidth]{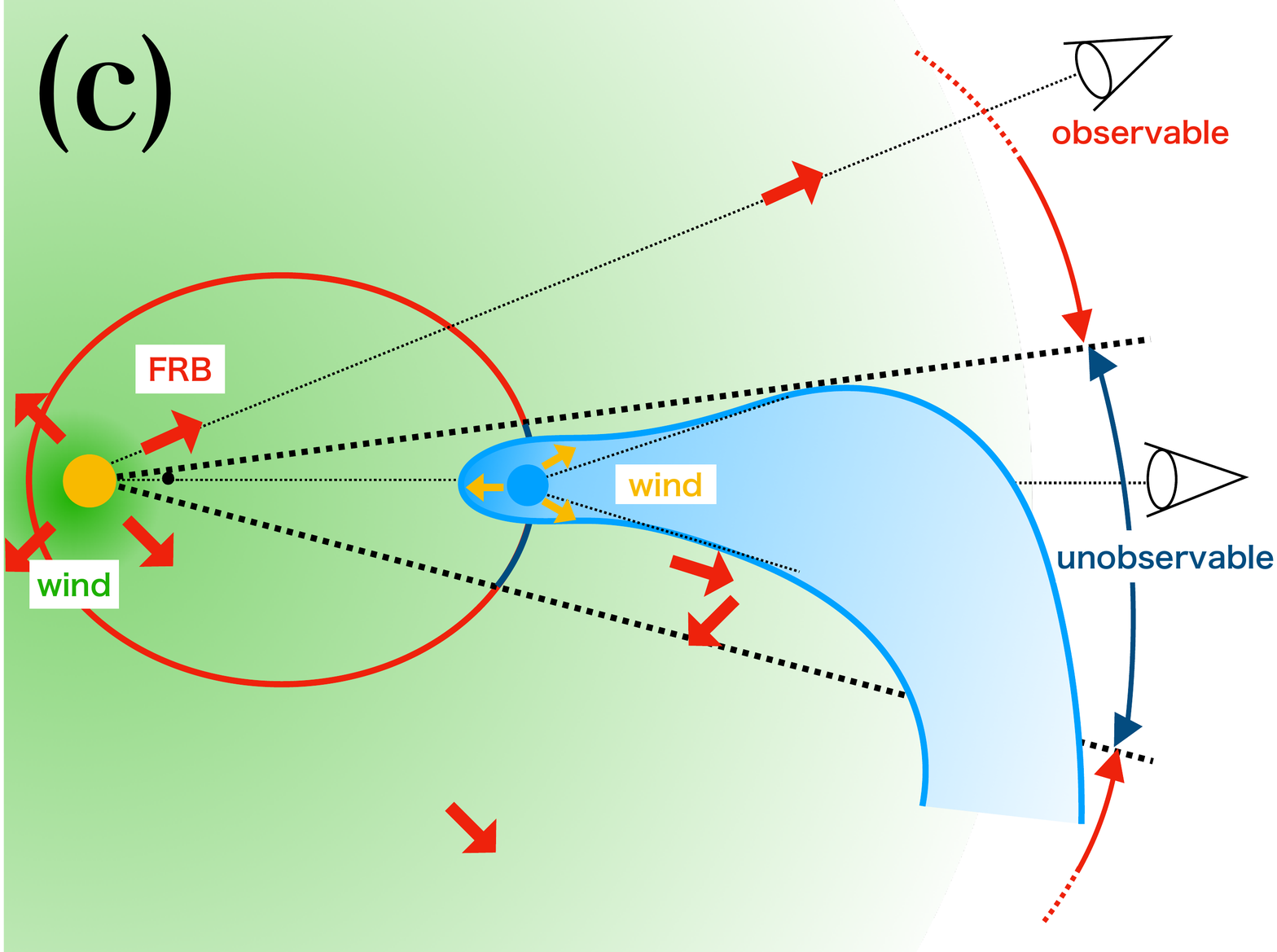}
          \caption{
          The three modes of the binary comb model:
          (a) The funnel mode for a periodic FRB.
          The yellow circle is the FRB pulsar, and the blue circle is the companion.
          The ellipsoid is the orbit of the FRB pulsar.
          Several lengths are shown: $R$ is the separation of the binary, $r_{\rm ph}$ is the photospheric radius, $r_{\rm s}$ is the length of the spiral arm, and $R_{\rm f}$ is the funnel size. The angle $\theta_{\rm c}$ is the half-opening angle of the funnel and $\theta_{\rm d}$ is the required angle to explain the duty cycle.
          The red curve corresponds to the active phase and the blue curve corresponds to the inactive phase for an observer on the right.
          In the funnel mode, $\theta_{\rm c}=\theta_{\rm d}$ and $r_{\rm s}>r_{\rm ph}>R_{\rm f}$ at $f=\pi\pm \theta_{\rm d}$ are required.
          (b) The $\tau$-crossing mode for a periodic FRB.
          $R_{\rm f}$ is the funnel size.
          The optical depth, $\tau$, is measured from infinity to $R_{\rm f}$.
          If the optical depth is lower than $10$ at $r=R_{\rm f}$, the FRB is observable.
          The optical depth changes depending on the orbital phase:
          $\tau(R_{\rm f})<10$ in the active phase (on the red curve), and $\tau(R_{\rm f})\geq10$ in the inactive phase (on the blue curve).
          In the $\tau$-crossing mode, $\theta_{\rm c}\leq\theta_{\rm d}$ and $r_{\rm ph}=R_{\rm f}$ at $f=\pi\pm \theta_{\rm d}$ are required.
          (c) The inverse funnel mode for a periodic FRB.
          In this figure, the ellipsoid is the orbit of the companion.
          The companion wind creates a funnel in the strong FRB pulsar wind.
          If this funnel is on the line of sight, the FRB is not observable (inactive phase).
          If the binary has non-zero eccentricity, the active window for the observers in the direction of the periapsis of the FRB pulsar is longer than that in the direction of the apoapsis.
          }
          \label{fig:funnel}
  \end{center}
\end{figure}

The binary comb model explains the periodic activity of the FRBs by the periodic variation of the optical depth along the line of sight.
There are three 
modes
of configurations for the opacity variation.
The first mode is the funnel mode (Fig.~\ref{fig:funnel}a) proposed by \cite{IokZha2020} and \cite{LyuBar2020}.
In the optically thick companion wind, the FRB pulsar wind would create a funnel.
The FRBs are observable only when the funnel points toward Earth because the opacity is low in the funnel.
The second mode is the $\tau$-crossing mode (Fig.~\ref{fig:funnel}b), which we find in this paper for the first time.
In this mode, the orbit intersects the photosphere and the FRB pulsar goes in and out the photosphere during the period.
The last mode is the inverse funnel mode (Fig.~\ref{fig:funnel}c), which is also identified in this paper for the first time.
In this mode, the FRB pulsar wind is stronger than that of the companion.
The companion wind points toward the opposite direction of the FRB pulsar and blocks the FRBs from the FRB pulsar, producing the periodicity.
Throughout this paper, we assume that the FRB pulsar wind does not absorb or scatter the FRBs.

In this section, we assume that the binary is observed from the direction of its periapsis in an edge-on configuration.
We also assume that the FRBs are radiated isotropically for simplicity.
The isotropic radiation does not necessarily mean that individual FRBs are radiated spherically.  Each FRB emission may be beamed and we assume that it is radiated in random direction.
The general viewing angle case and the effect of beaming are discussed in Sec.~\ref{sec:121102} for FRB 121102 and in Appendix~\ref{sec:general_angle}.

\subsection{funnel mode}\label{sec:funnel}
In the funnel mode, FRBs are observable when the funnel points toward Earth because the funnel created by the FRB pulsar wind makes the optical depth smaller.
Thus, the observed duty cycle determines the half-opening angle of the funnel.
Also, the funnel must extend in the radial direction to the outside of the photosphere, otherwise the FRBs are eventually blocked by the companion wind.
To evaluate these conditions, we calculate three quantities; the range of the true anomaly where the FRBs are observable, the half-opening angle of the funnel, and the length of the spiral arm where the tail of the funnel is spiraled by the orbital motion.

\begin{figure}[H]
\begin{center}
          \includegraphics[clip, width=0.6\textwidth]{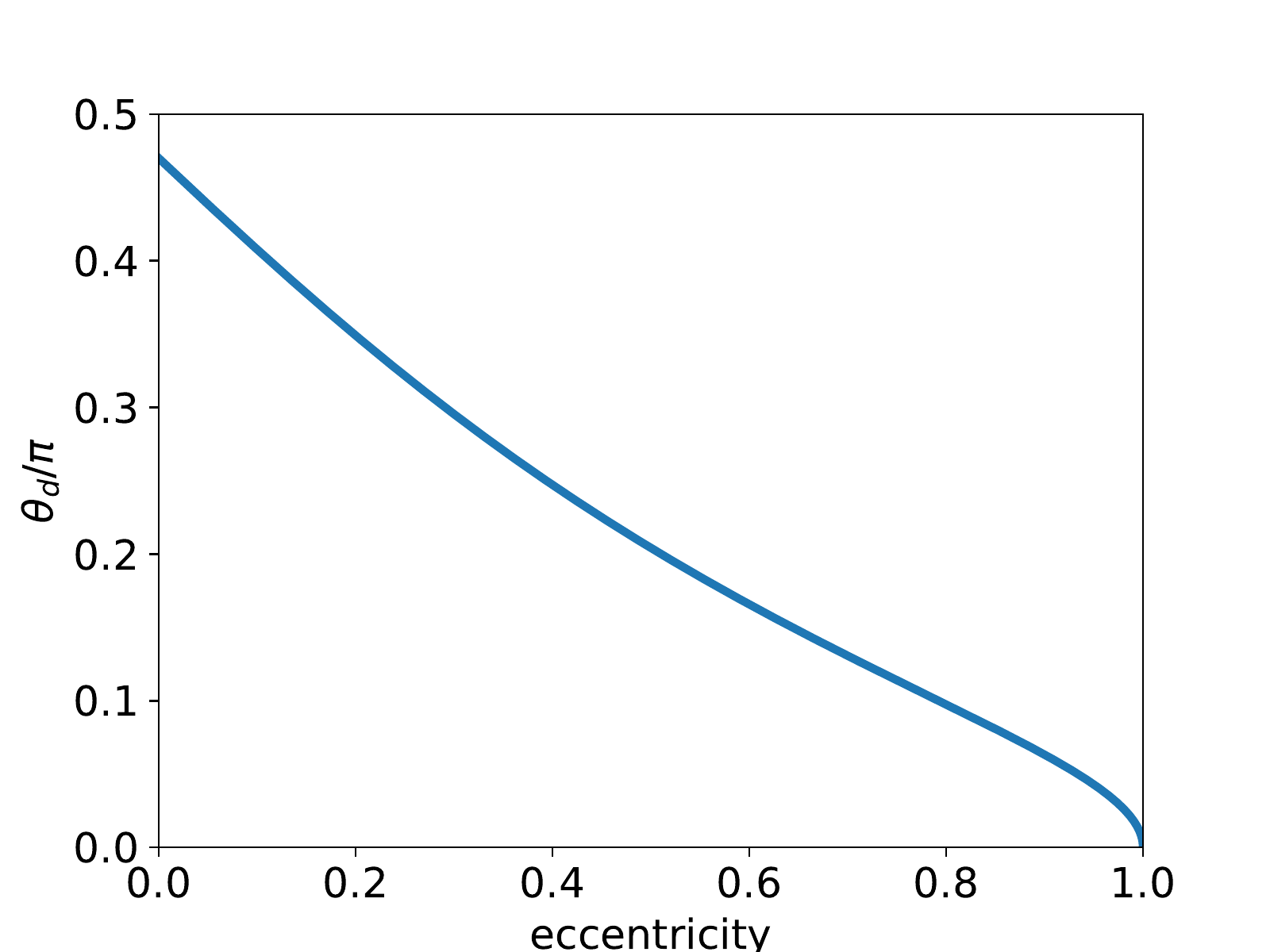}
          \caption{
            The relation between the eccentricity, $e$, and $\theta_{\rm d}$, the required angle to explain the duty cycle.
            The duty cycle of the periodic FRB equals 47\% in this figure.
            This relation is derived by solving Eq.~(\ref{eq:duty_edgeon}).
            }
              \label{fig:ecc_thetac}
  \end{center}
\end{figure}

For a given duty cycle, the range of the true anomaly where the FRBs are observable is determined as follows.
As we can generally take the coordinates so that $f$ increases with time, the FRB becomes observable when the true anomaly, $f$, equals $\pi-\theta_{\rm d}$ and becomes unobservable when $f=\pi+\theta_{\rm d}$.
The time taken to orbit from $f=\pi-\theta_{\rm d}$ to $f=\pi+\theta_{\rm d}$ is expressed as \citep{PoissonWill2014}
\be
T=2\int_{\pi-\theta_{\rm d}}^\pi\sqrt{\frac{(1-e^2)^3a^3}{GM_{\rm tot}}}\frac{df}{(1+e\cos f)^2}.
\ee
With the eccentric anomaly $u$, this integral is expressed only by elementary functions as
\be
T=2\sqrt{\frac{a^3}{GM_{\rm tot}}}\lsb\pi-\lrb u_d-e\sin u_d\rrb\rsb,
\ee
where $u_d=2\tan^{-1}\lrb\sqrt{\frac{1-e}{1+e}}\tan\frac{\pi-\theta_{\rm d}}{2}\rrb$.
To realize the observed duty cycle, $\sim47$\%, $T/P$ must be equal to 0.47 where $P=2\pi\sqrt{\frac{a^3}{GM_{\rm tot}}}$ is the orbital period.
This condition is given as 
\be
\frac{T}{P}=\frac{1}{\pi}\lrb \pi- u_d+e\sin u_d\rrb=0.47,
\label{eq:duty_edgeon}
\ee
which gives a relation between $e$ and $\theta_{\rm d}$.
Figure~\ref{fig:ecc_thetac} shows this relation.
The higher the eccentricity, the smaller the angle $\theta_{\rm d}$.
This is because the FRB pulsar spends more time near the apoapsis for a higher eccentricity.

The half-opening angle of the funnel, $\theta_{\rm c}$, is determined by the pressure balance between the FRB pulsar wind and the companion wind.
Using the thin-shell approximation, $\theta_{\rm c}$ is determined by
\be
\begin{aligned}
-\theta_{c}+\tan \theta_{c}&=&\frac{\beta\pi}{1-\beta},\\
\beta&=&\frac{L_{\rm w} / c}{\dot{M} V},
\end{aligned}
\ee
where $L_{\rm w}$ is the luminosity of the FRB pulsar wind \citep{CanRag1996}.
We consider the case that $\beta$ is smaller than unity in the funnel mode.\footnote{
The case that $\beta\geq 1$ is considered in the inverse funnel mode (Sec.~\ref{sec:inverse}).}
$\theta_{\rm c}$ does not depend on the true anomaly or the semi-major axis of the binary and thus does not change with time.
Here, we neglect the effect of the orbital motion on the half-opening angle because the orbital motion is much slower than the stellar wind.
We assume that the shape of the funnel is a cone whose vertex is the FRB pulsar and whose half-opening angle equals $\theta_{\rm c}$.
The FRBs are observable while the line of sight is inside the cone.
The opening angle of the funnel becomes almost constant above a radius $R_{\rm f}$ (see Fig.~\ref{fig:funnel}) where the pressure of the companion wind balances the FRB pulsar wind.
This radius, $R_{\rm f}$, which we call the funnel size, is determined by
\be
\frac{\dot{M}}{4\pi V(R^2+R_{\rm f}^2)}V^2\sim \frac{L_{\rm w}}{4\pi cR_{\rm f}^2},
\label{eq:rf}
\ee
where we have used the approximation that $R^2+R_{\rm f}^2+2RR_{\rm f}\cos\theta\sim R^2+R_{\rm f}^2$.
Solving Eq.~(\ref{eq:rf}), we obtain
\be
R_{\rm f}\sim\sqrt{\frac{1}{\beta^{-1}-1}}R.
\ee
For a high $\beta$ value, one has $R_{\rm f}\sim\sqrt{1/(1-\beta)}R$, and for a low $\beta$ value, one has $R_{\rm f}\sim\sqrt{\beta}R$.
We adopt $R_{\rm f}$ to evaluate the optical depths (see Sec.~\ref{sec:opticallythick}).
Although the exact funnel size depends on the direction of the observer, we adopt $R_{\rm f}$ as the typical value of funnel size and do not consider its angular structure.

\begin{figure}[H]
  \begin{center}
          \includegraphics[clip, width=0.7\textwidth]{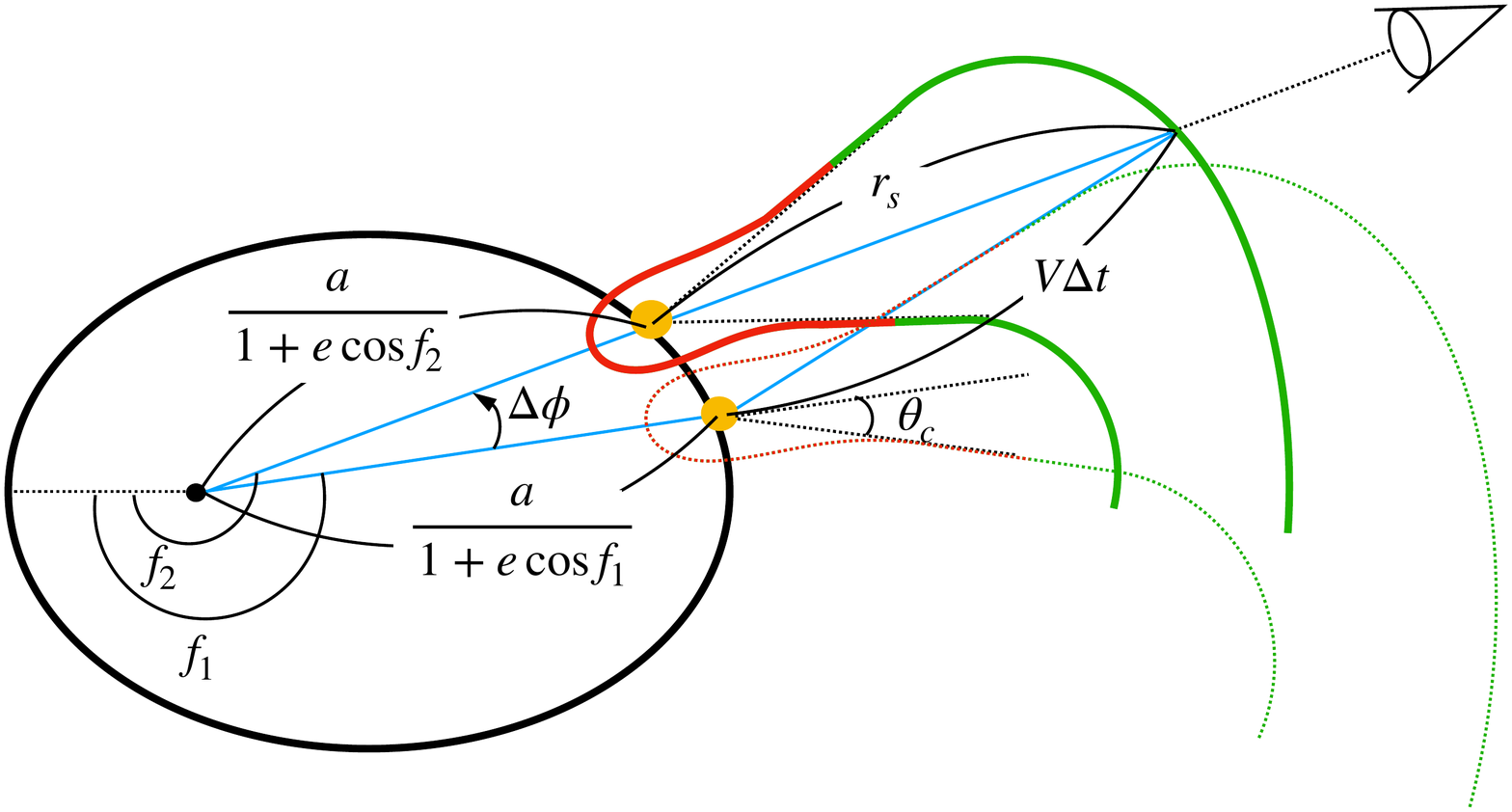}
          \caption{
            The configuration of the funnel used to calculate the length of the spiral arm $r_{\rm s}$.
            The sine theorem is applied to the blue triangle.}
        \label{fig:rs}
  \end{center}
\end{figure}

In the funnel mode, the length of the spiral arm, $r_{\rm s}$, where the tail of the funnel is spiraled by the orbital motion \citep{ParPit2008, BosBar2015}, should be larger than the photospheric radius, $r_{\rm ph}$.
We evaluate $r_{\rm s}$ by splitting the contact discontinuity of the funnel into two sections.
One is a shocked cap where the orbital motion does not affect the structure of the bow shock (the red curve in Fig.~\ref{fig:rs}).
The other is the ballistic contact discontinuity where the orbital motion changes the shape of the funnel (the green curve in Fig.~\ref{fig:rs}).
In the ballistic contact discontinuity, the fluid elements flow ballistically with the velocity $V$ because the densities of both winds have almost the same dependence on distance.
For simplicity, we ignore the skew of the shock cap by the orbital motion discussed in \cite{ParPit2008} and assume that the fluid element in the ballistic contact discontinuity has a half-opening angle $\theta_{\rm c}$.
To calculate $r_{\rm s}$, we introduce two true anomalies $f_1,\,f_2$, and a time duration $\Delta t$. Here $f_2$ is the true anomaly when the FRB pulsar is on the line of sight from the observer to the companion,
$f_1$ is the true anomaly such that the fluid element ejected at $f_1$ along the contact discontinuity reaches the line of sight at $f_2$ (see Fig.~\ref{fig:rs}), and
$\Delta t$ is the time duration taken for the FRB pulsar to orbit from $f_1$ to $f_2$.
Using the sine theorem, we obtain the relation between $r_{\rm s}$ and the other parameters as 
\be
\frac{a(1+e\cos f_2)^{-1}+r_{\rm s}}{\sin\theta_{\rm c}}=\frac{V\Delta t}{\sin \lrb f_2-f_1\rrb}=\frac{a(1+e\cos f_1)^{-1}}{\sin\left(\theta_{\rm c}-(f_2-f_1)\right)}.
\ee
For a given $f_2$, we obtain $f_1$ using the second equation and by substituting $f_1$ into the first equation, we obtain $r_{\rm s}$.
Assuming that $|f_1-f_2|\ll1$ and thus $f_2-f_1\simeq\dot{\phi}\Delta t$, we obtain the length of the spiral arm as 
\bea
r_{s}&=&\sin \theta_{c} \frac{V}{\dot{\phi}} - \frac{a}{1+e \cos f_2}\nn\\
&=&\sin\theta_{\rm c}\frac{V}{\Omega}\lrb1-e^2\rrb^{3/2}\lrb1+e\cos f_2\rrb^{-2}- \frac{a}{1+e \cos f_2}, 
\label{eq:r_s}
\eea
where $\Omega=\sqrt{\frac{GM_{\rm tot}}{a^3}}$ and $\dot{\phi}=\Omega \lrb1-e^2\rrb^{-3/2}\lrb1+e\cos f_2\rrb^{2}$.

There are three conditions for the funnel mode.
The first condition is that the companion wind must be optically thick to FRBs so that the FRBs in the inactive phase are absorbed or scattered by the companion wind ($r_{\rm ph}>R_{\rm f}$).
The second  condition is that the half-opening angle of the funnel must be such that the duty cycle can be explained ($\theta_{\rm c}=\theta_{\rm d}$).
The third condition is that the funnel must extend radially enough for FRBs not to be blocked by the companion wind.
Thus, the length of the spiral arm must be larger than the photospheric radius ($r_{\rm s}>r_{\rm ph}$).
These conditions for the funnel mode are represented as
\bea
\theta_{\rm c}=\theta_{\rm d},\label{eq:funnel1}\\
r_{\rm s}>r_{\rm ph}>R_{\rm f}\label{eq:funnel2}.
\eea
The latter condition is evaluated at $f=\pi\pm\theta_{\rm d}$.

\subsection{$\tau$-crossing mode}\label{sec:tau}
In the $\tau$-crossing mode, the FRBs in the active phase become observable because the optical depth on the line of sight gets lower than $10$ (Fig.~\ref{fig:funnel}b).
This condition requires that the photospheric radius equals the funnel size at $f=\pi\pm\theta_{\rm d}$ where $\theta_{\rm d}$ is obtained by solving Eq.~(\ref{eq:duty_edgeon}).
Also, the half-opening angle of the funnel must be small enough that the FRBs in the inactive phase do not become observable through the funnel.
This condition requires $\theta_{\rm c}\leq\theta_{\rm d}$.
Thus, the conditions for the $\tau$-crossing mode are
\bea
\theta_{\rm c}\leq\theta_{\rm d},\label{eq:tau1}\\
r_{\rm ph}=R_{\rm f}\label{eq:tau2}.
\eea
The latter condition is evaluated at $f=\pi\pm\theta_{\rm d}$.

\subsection{Inverse funnel mode}\label{sec:inverse}
In this subsection, we consider the case that $\beta=L_{\rm w}/\dot{M}Vc\geq 1$.
In this case, the companion wind is pushed back by the FRB pulsar wind and points to the inverse direction of the FRB pulsar (Fig.~\ref{fig:funnel}c).
Still the companion wind may block FRBs from the FRB pulsar.
We approximate the shape of the companion wind as a cone whose half-opening angle equals $\theta_{\rm c}$ (see Sec.~\ref{sec:funnel}).

In the case of $\beta\geq 1$, for any $L_{\rm w}$ and $\dot{M}$, the observed duty cycle is realized by choosing viewing angle and eccentricity appropriately.
The reason is as follows.
We first consider that the binary is observed  edge-on.
For given $L_{\rm w}$ and $\dot{M}$, the funnel shape is determined.
In the inverse funnel mode, FRBs are not observable as the line of sight is inside the cone of the companion wind.
Let us assume $e=0$.
Then, the duty cycle of the periodic FRB is determined uniquely.
If this duty cycle is shorter than the observed value, by letting the eccentricity higher and observing the binary from the direction of the periapsis of the FRB pulsar, the duty cycle gets longer and the observed duty cycle is realized.
In contrast to this, if the duty cycle for $e=0$ is longer than the observation, by increasing  eccentricity and observing from the direction of the apoapsis of the FRB pulsar, the duty cycle gets shorter and the observed duty cycle is realized.
Changing the inclination angle increases the duty cycle because the FRB pulsar is more easily observed.

This mode differs from the case of $\beta<1$ in the following way.
The FRB pulsar wind is stronger than that of the companion in this mode.
Thus, the companion wind cannot affect the magnetosphere of the FRB pulsar and the FRB emission is not affected by the binary interaction.
As a result, one of the models to realize the frequency-dependence of the active window is not applicable to the case of $\beta\geq 1$
because any aurora particles are not expected
(see Sec.~\ref{sec:frequency} for more details).

In this mode, only the constraint $\beta=L_{\rm w}/\dot{M}Vc>1$ is imposed on $L_{\rm w}$ and $\dot{M}$.
Thus we do not consider this mode later.
We mainly consider the case of $\beta<1$.

\section{Other constraints}\label{sec:otherconstraint}
\subsection{Change in dispersion measure}\label{sec:dDM}
The observed change in the dispersion measure, ${\Delta \rm DM}$, also gives constraint on the host binary.
The change of DM due to orbital motion must be smaller than the observed upper limit of variability, which is of the order $6\,{\rm cm^{-3}}\,{\rm pc}$ for FRB 121102 \citep{Li2021_FAST}.
We calculate the dispersion measure, ${\rm DM}$, from the edge of the spiral arm, $r_{\rm s}$, to infinity.
$\Delta {\rm DM}$ is obtained by comparing the values at $f=\pi$ (apoapsis) and at $f=\pi+\theta_{\rm d}$.
In the calculation, we ignore the $\theta$-dependence of the wind number density and set $\theta=0$ for simplicity.
The dispersion measure is given by
\bea
{\rm DM} &\simeq& \int_{r_{\rm s}}^{\infty} n d r\nn\\
&=&\int_{r_{\rm s}}^{\infty} \frac{\dot{M}}{4 \pi m_{p} V \left(r^2+R^2+2Rr\right)} dr\nn\\
&=&\frac{\dot{M}}{4\pi m_pV}\frac{1}{r_{\rm s}+R}\nn\\
&=&3.24\times10^{7}\,{\rm cm^{-3}pc}\, \frac{1}{r_{\rm s}/R+1}\lrb\frac{\dot{M}}{M_\odot{\rm yr}^{-1}}\rrb\lrb\frac{R}{10^{14}\,{\rm cm}}\rrb^{-1}\lrb\frac{V}{0.01c}\rrb^{-1}.
\label{eq:DM}
\eea
Thus, the difference of ${\rm DM}$ in the active phase is given by
\bea
\Delta {\rm DM}&=&\frac{\dot{M}}{4\pi m_pV}\left(\left.\frac{1}{r_{\rm s}+R}\right|_{f=\pi}-\left.\frac{1}{r_{\rm s}+R}\right|_{f=\pi+\theta_{\rm d}}\right)\nn\\
&=&\frac{\dot{M}}{4\pi m_pV a X(\pi)}\left[1-\frac{X(\pi)}{X(\pi+\theta_{\rm d})}\right]\label{eq:dDM},
\eea
where we have defined 
\be
X(f)=\left.\frac{r_{\rm s}+R}{a}\right|_{f}=\sin\theta_{\rm c}\frac{V}{a\Omega}(1-e^2)^{3/2}(1+e\cos(f))^{-2}+\frac{M_{\rm NS}}{M_c}\frac{1}{1+e\cos{f}},
\label{eq:X}
\ee
and used Eq.~(\ref{eq:r_s}).
This value must be smaller than the observed upper limit on the change of DM.

We note that the formula Eq.~(\ref{eq:dDM}) is also used for the $\tau$-crossing mode.
In the $\tau$-crossing mode, this is a conservative constraint on the $\Delta {\rm DM}$.
If $\theta_{\rm c}<\theta_{\rm d}$, FRBs enter the companion wind at $\sim R_{\rm f}$ instead of $\sim r_{\rm s}$ and $\Delta {\rm DM}$ becomes higher than Eq.~(\ref{eq:dDM}).

\subsection{Persistent radio  counterpart}\label{sec:persistent}
We consider the consistency with the persistent radio counterpart of the FRB 121102
\citep{Cha2017,Mar2017}.
There are two possibilities for the source of this persistent radio counterpart.

One is the synchrotron nebula emission powered by the pulsar wind of the FRB pulsar or by the magnetar flare.
The nebula is formed by the interaction between the supernova ejecta and the FRB pulsar wind \citep{KasMur2017,YanZha2016}, by the interaction between the interstellar matter and the FRB pulsar wind\footnote{In this case, the FRB pulsar must have a high kick velocity.} \citep{DaiWan2017}, or by the interaction between the supernova ejecta and the ejecta of the magnetar flare \citep{Bel2017}.
In these models, the wind luminosity of the FRB pulsar, $L_{\rm w}$, should be high enough to energize the persistent radio counterpart.
If the persistent radio counterpart is due to the nebula emission,
\be
L_{\rm w}\geq L_{\rm w,lower}
\label{eq:persistent_nebula}
\ee
is required where $L_{\rm w, lower}$ is the lower limit on the wind luminosity of the FRB pulsar to energize the persistent radio counterpart.
We note that the value depends on the parameters of the nebula, such as the age of the nebula, the ejecta mass, the energy of the supernova, and so on.
For FRB 121102, we adopt $L_{\rm w, lower}\sim 10^{40}\,{\rm erg\, s^{-1}}$ \citep{Bel2017,KasMur2017,YanZha2016,DaiWan2017}.

The other possibility is that the persistent radio counterpart is due to the activity of the companion.
For example, for the SMBH companion case, the persistent radio counterpart might be due to the AGN activity of the SMBH \citep{Zha2018}.
The kinetic luminosity of the outflow equals $\frac{1}{2}\dot{M}V^2$ and some fraction of this luminosity would be converted into the radio luminosity, $L_{\rm radio}$.
Thus, in this case, the kinetic luminosity must be higher than the luminosity of the persistent radio counterpart,
\be
\frac{1}{2}\dot{M}V^2> L_{\rm radio}.
\label{eq:persistent_companion}
\ee
This is the least necessary condition for this case because the emission efficiency in the radio band would be less than unity.

The radio efficiency would change the constraint of Eq.~(\ref{eq:persistent_companion}).
If an $\epsilon_{\rm e}$ fraction of the kinetic energy is converted to electron energy and an $\eta_{\rm r}$ fraction of the electron energy is radiated in the radio band, we obtain
\be
\epsilon_{\rm e}\eta_{\rm r}\,\frac{1}{2}\dot{M}V^2\geq L_{\rm radio}
\ee
instead of Eq.~(\ref{eq:persistent_companion}).
We adopt $\epsilon_{\rm e}\eta_{\rm r}\sim10^{-3}$ in later discussions.

In the SMBH or IMBH case, the radio emission may originate from synchrotron radiation
in the jet.
In that case, some fraction of the accretion luminosity, $\dot{M}_{\rm acc}c^2$ where $\dot{M}_{\rm acc}$ is the  mass accretion rate of the black hole, is converted to the total jet luminosity, and some of the jet luminosity is radiated in the radio band.
On the other hand, some fraction of the accretion luminosity, $\dot{M}_{\rm acc}c^2$, is converted to the outflow kinetic luminosity, $\dot{M}V^2/2$.
Thus, the $\dot{M}V^2$ factor may not be directly related to $L_{\rm radio}$ so that conditions looser than Eq.~(\ref{eq:persistent_companion}) is possible.

\section{Binary comb model for FRB 121102}\label{sec:121102}
Using the binary comb model discussed so far, we constrain the parameters of the host binary of FRB 121102.
The parameters used here are as follows.
The period of FRB 121102
is $159$ days and its duty cycle
is $47$\% \citep{Raj2020,Cru2020} (The frequency dependence of the duty cycle is discussed in Sec.~\ref{sec:frequency}).
The change in the dispersion measure, $\Delta{\rm DM}$, 
is less than
$6\,{\rm cm^{-3}}\,{\rm pc}$ \citep{Li2021_FAST}.
Also, FRB 121102 is associated with the persistent radio counterpart whose luminosity, $L_{\rm radio}$, 
is
$10^{39}\,{\rm erg\,s^{-1}}$ \citep{Cha2017,Mar2017}.
If this persistent radio counterpart is the emission of a synchrotron nebula, the lower limit on the pulsar wind luminosity of the FRB pulsar 
is adopted as
$L_{\rm w, lower}= 10^{40}\,{\rm erg\, s^{-1}}$ (see Sec.~\ref{sec:persistent}).
We fix the frequency of FRBs, $\nu$, as $1\,{\rm GHz}$, the energy of FRBs, $L_{\rm FRB}\Delta t$, as $1\times10^{38}\,{\rm erg}$, and the temperature of the wind, $T$, as $10^4\,{\rm K}$.
The FRB energy is adopted as $L_{\rm FRB}\Delta t=1\times10^{38}\,{\rm erg}$, at which the event rate is highest in \cite{Cru2020}.

There are seven parameters for the binary, $M_{\rm c}$, $V$, $e$, $\dot{M}$, $L_{\rm w}$, the viewing angle, and the inclination angle.
First, we consider the case that the binary is observed from the direction of its periapsis with an edge-on geometry.
The effects of the viewing angle and the inclination angle are discussed later in this section
and in Appendix~\ref{sec:general_angle}.
In this section, for given $M_{\rm c}$ and $V$ values, we put constraints on $\dot{M}$, $L_{\rm w}$, and $e$.
As in the previous section, we assume that each FRB is radiated spherically or the FRBs are radiated in random directions.
The beaming effect is considered later.

\begin{figure}[H]
  \begin{center}
    \begin{tabular}{c}
      \begin{minipage}{0.5\hsize}
        \begin{center}
          \includegraphics[clip,width=\textwidth]{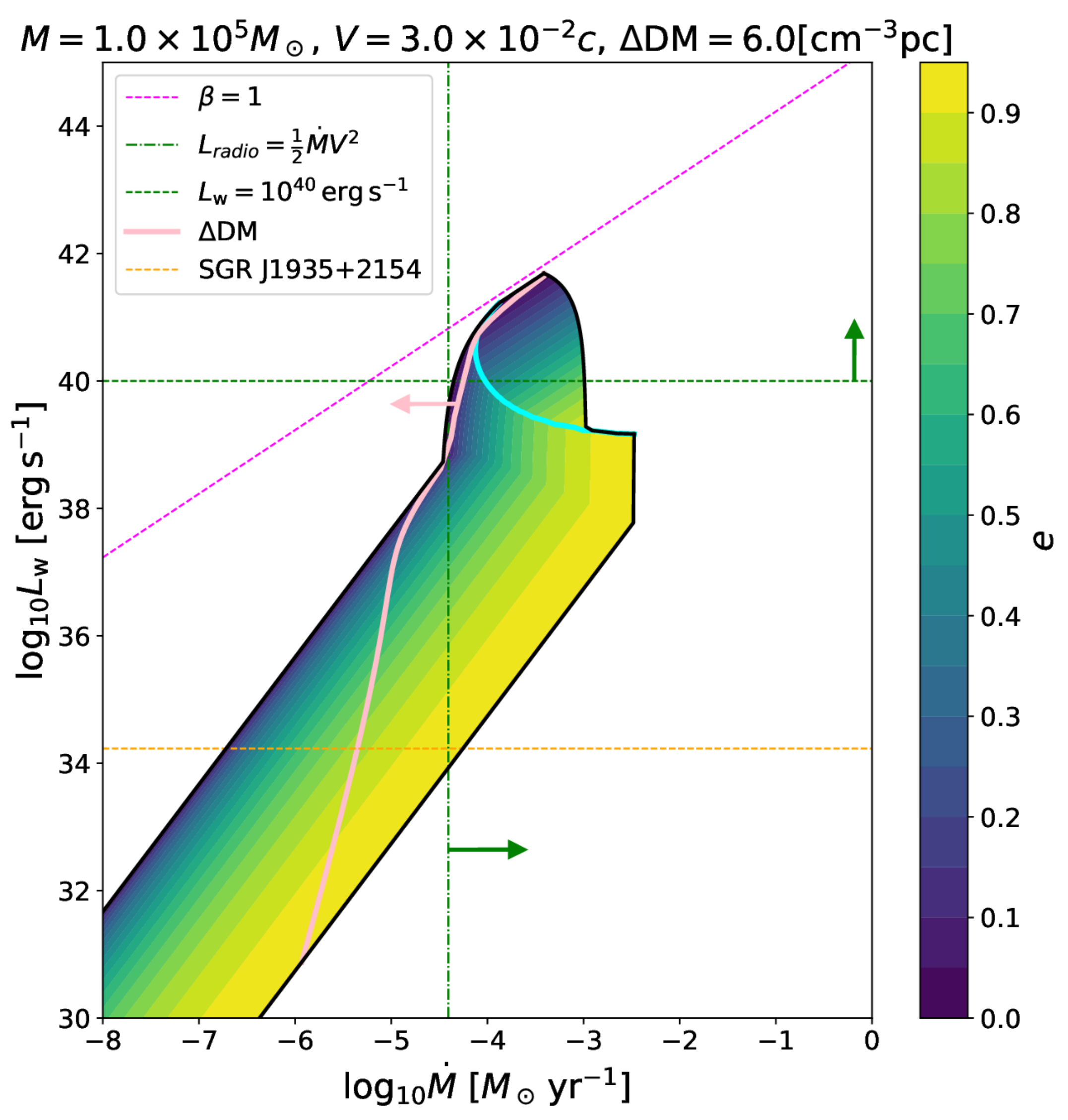}
        \end{center}
      \end{minipage}
      \begin{minipage}{0.5\hsize}
        \begin{center}
          \includegraphics[width=\textwidth]{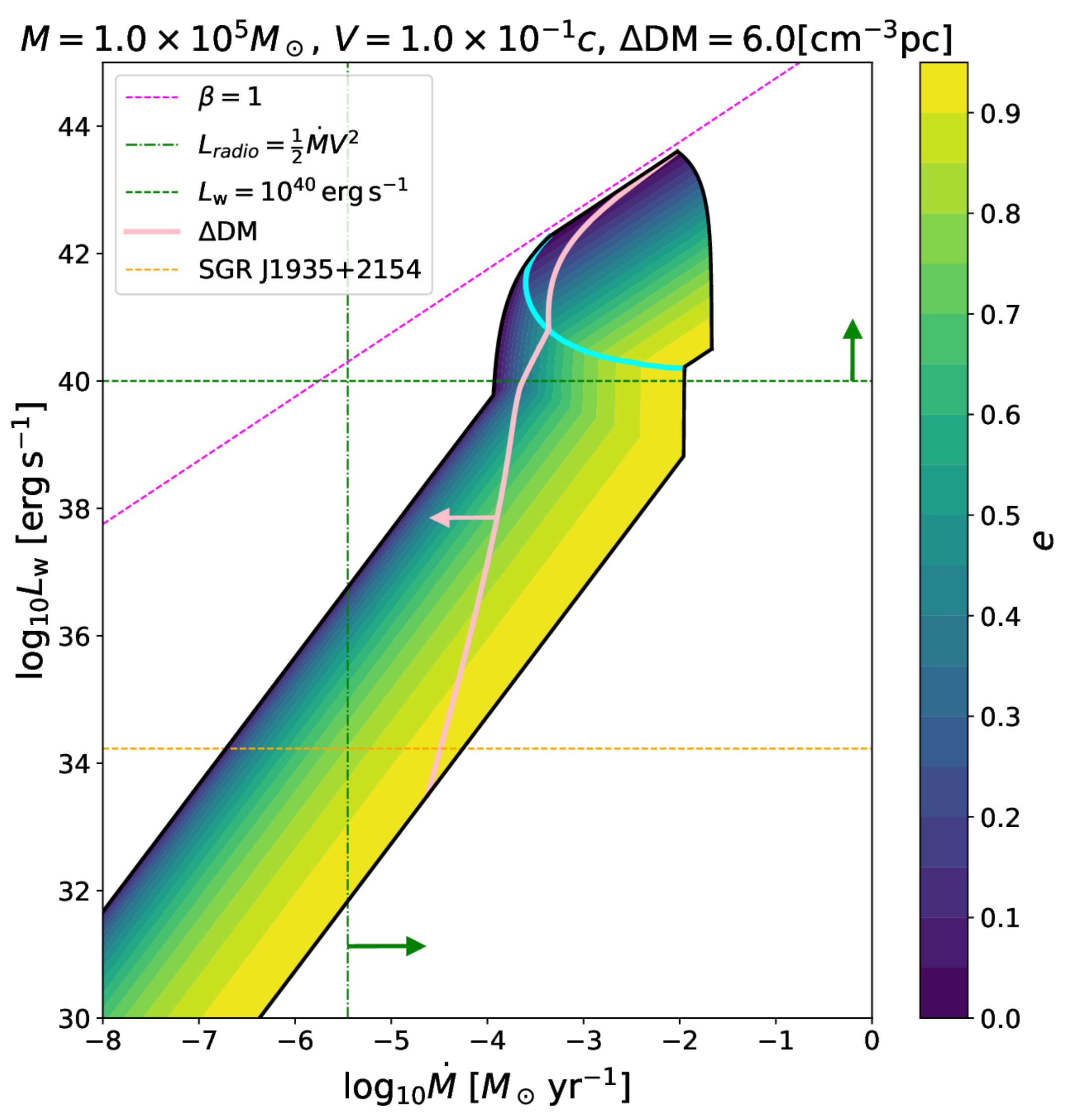}
        \end{center}
      \end{minipage}
    \end{tabular}
    \caption{
    The allowed region in the $\dot{M}$-$L_{\rm w}$ plane for the SMBH ($10^5M_\odot$) case where $\dot{M}$ is the mass-loss rate of the companion and $L_{\rm w} $ is the wind luminosity of the FRB pulsar.
    The binary is observed from the direction of its periapsis with an edge-on geometry.
    The left panel shows the $V=3\times10^{-2}c$ case and the right panel shows the $V=1\times10^{-1}c$ case, where $V$ is the velocity of the wind of the SMBH.
    The colored region is allowed by the funnel mode or the $\tau$-crossing mode, and the color contour shows the lower limit on eccentricity.
    The region above the magenta dashed line, where the companion wind is pushed back by the FRB pulsar wind, is allowed by the inverse funnel mode.
    The light pink line shows the line where $\Delta{\rm DM}=6\,{\rm cm^{-3}\,pc}$, and the right side of this line is excluded.
    Above and below the cyan line in the colored region, the funnel mode and the $\tau$-crossing mode are applied, respectively.
    The left edge of the funnel mode (the cyan line) is determined by the condition that the binary system is optically thick.
    The right edge of the funnel mode is determined by the condition that the length of the spiral arm is longer than the photospheric radius.
    The persistent radio counterpart is powered either by the companion (the right side of the green dash-dotted line) or by the FRB pulsar (above the green dashed line).
    We note that if the radio efficiency is $\sim10^{-3}$, the green dash-dotted line moves three digits to the right.
    For reference, the orange line shows the spin-down luminosity of SGR J1935+2154, which produced FRB 200428.
    }
    \label{fig:allowed_SMBH}
  \end{center}
\end{figure}

\begin{figure}[H]
  \begin{center}
    \begin{tabular}{c}
      \begin{minipage}{0.5\hsize}
        \begin{center}
          \includegraphics[clip,width=\textwidth]{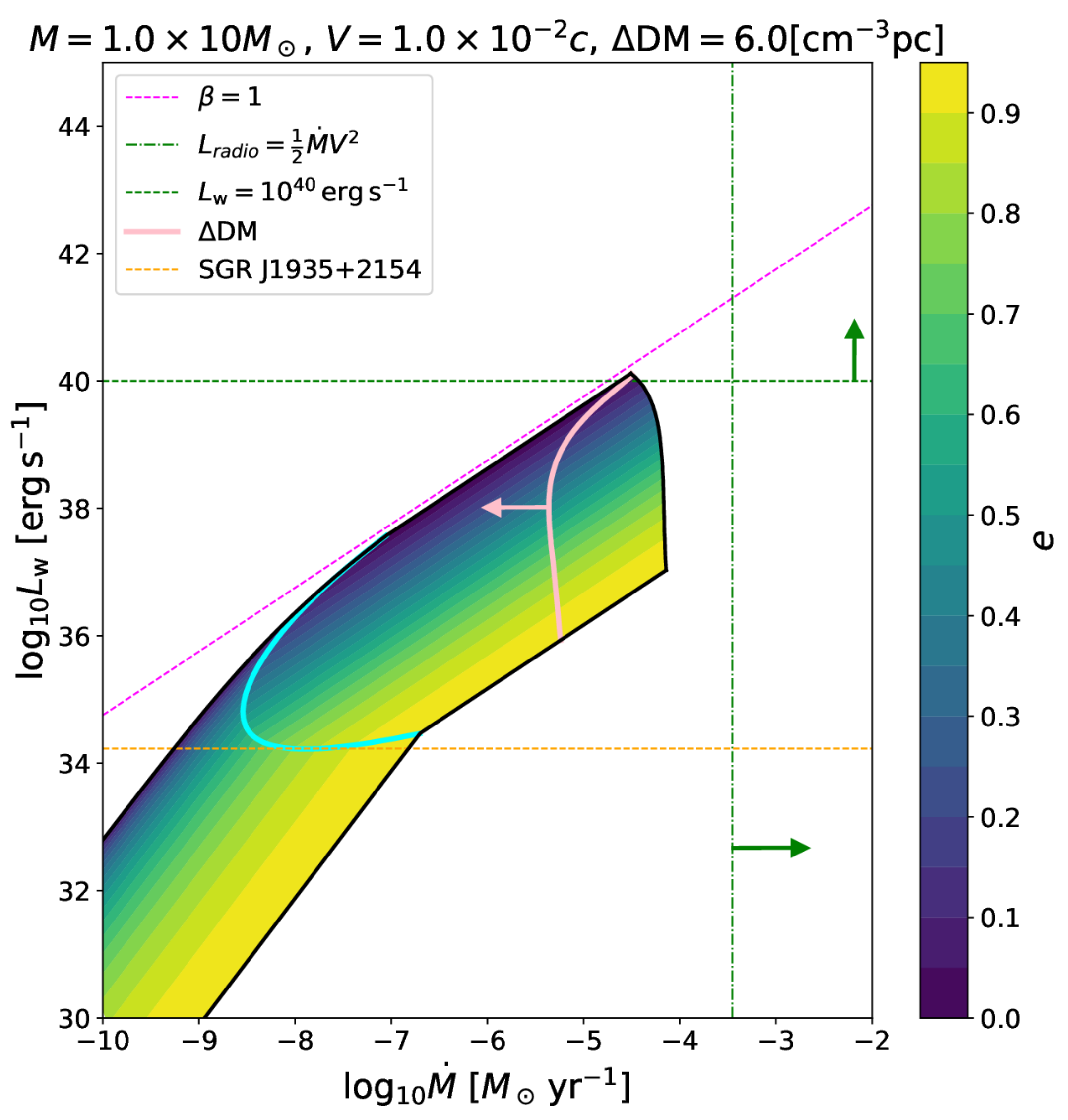}
        \end{center}
      \end{minipage}
      \begin{minipage}{0.5\hsize}
        \begin{center}
          \includegraphics[width=\textwidth]{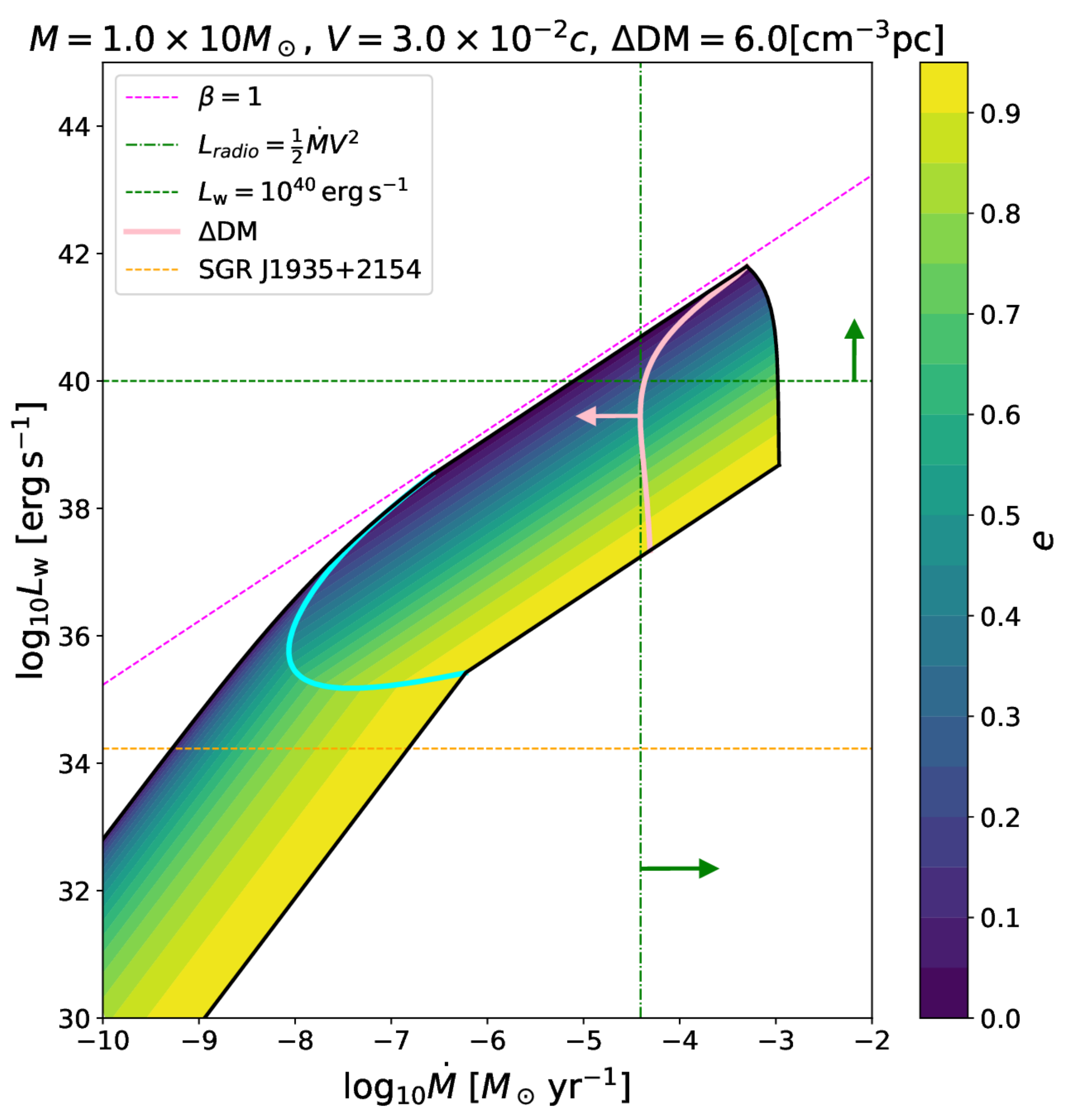}
        \end{center}
      \end{minipage}
    \end{tabular}
    \caption{The allowed region in the $\dot{M}$-$L_{\rm w}$ plane for the MS ($10M_\odot$) companion case.
      The left panel shows the $V=1\times10^{-2}c$ case and the right panel shows the $V=3\times10^{-2}c$ case. Other notations are the same as Fig.~\ref{fig:allowed_SMBH}.
      Almost all the regions are excluded in the case of $V=1\times10^{-2}c$, but there is an allowed region in the extremely high velocity case with $V=3\times10^{-2}c$.
            }
    \label{fig:allowed_MS}
  \end{center}
\end{figure}

\begin{figure}[H]
  \begin{center}
    \begin{tabular}{c}
      \begin{minipage}{0.5\hsize}
        \begin{center}
          \includegraphics[clip,width=\textwidth]{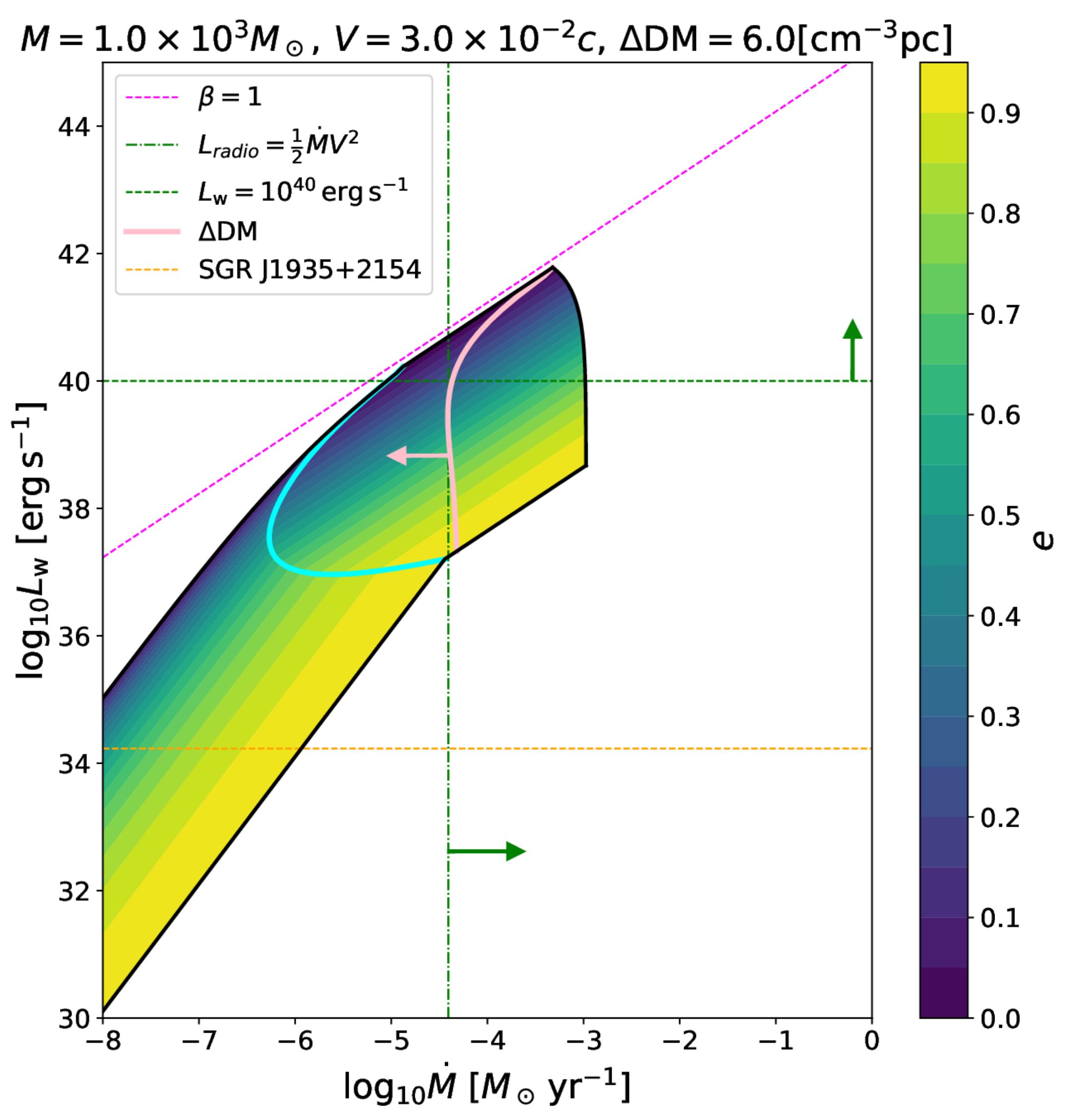}
        \end{center}
      \end{minipage}
      \begin{minipage}{0.5\hsize}
        \begin{center}
          \includegraphics[width=\textwidth]{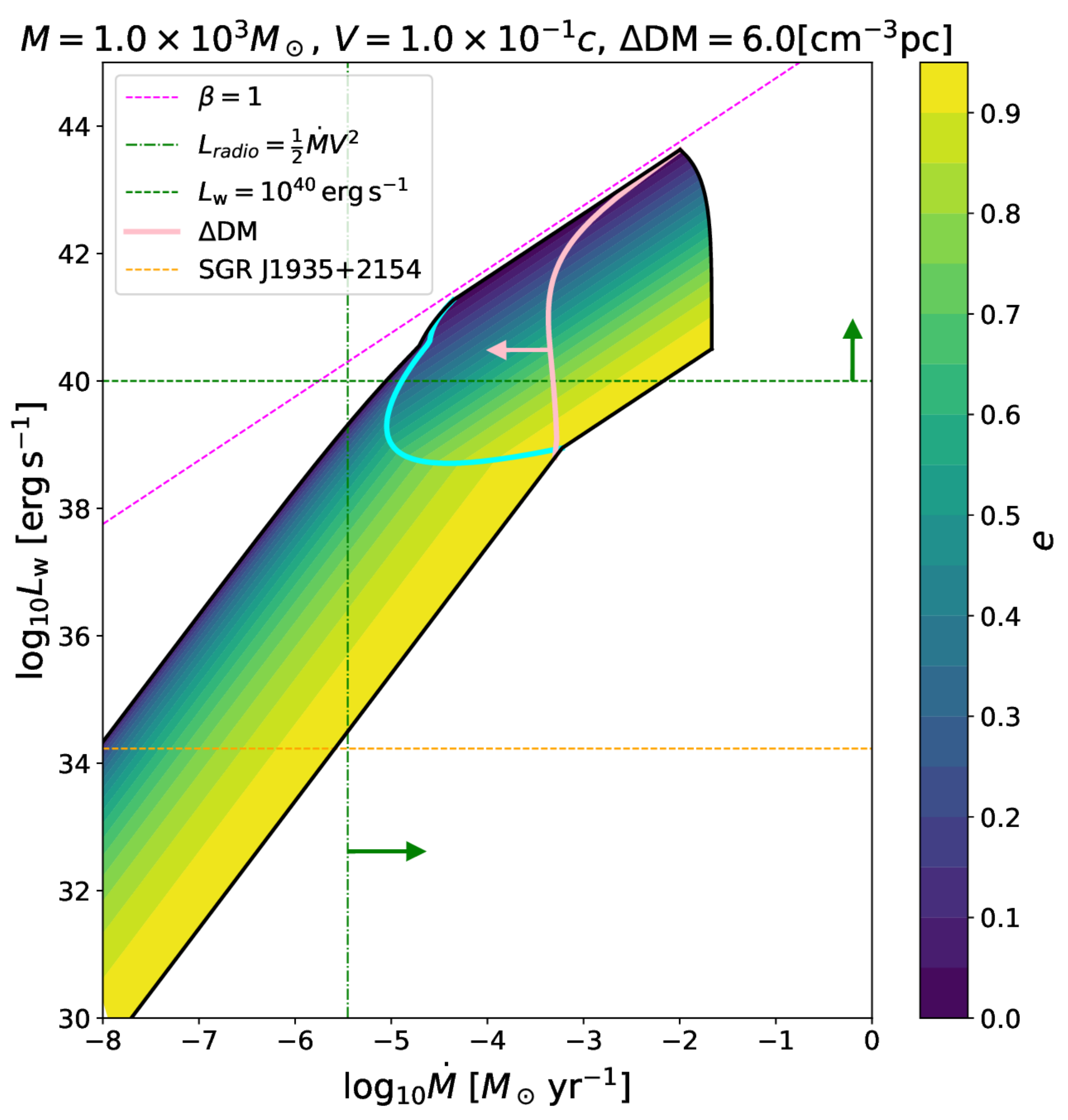}
        \end{center}
      \end{minipage}
    \end{tabular}
    \caption{The allowed region in the $\dot{M}$-$L_{\rm w}$ plane for the IMBH case.
      The left panel shows the $V=3\times10^{-2}c$ case and the right panel shows the $V=1\times10^{-1}c$ case.
      Other notations are the same as Fig.~\ref{fig:allowed_SMBH}.
      There are allowed regions in both cases.
      }      
    \label{fig:allowed_IMBH}
  \end{center}
\end{figure}

Figures \ref{fig:allowed_SMBH}--\ref{fig:allowed_IMBH} show the constraints on the binary system of FRB 121102 in the cases that the companion is an SMBH ($10^5M_\odot$), an MS ($10M_\odot$), and an IMBH ($10^3M_\odot$).
The colored region is allowed in the funnel mode or the $\tau$-crossing mode and the color contour shows the 
lower limit on the
eccentricity.
The cyan line represents the boundary between the funnel mode (above the line; Sec.~\ref{sec:funnel}) and the $\tau$-crossing mode (below the line; Sec.~\ref{sec:tau}).
For a given eccentricity, the region allowed by the funnel mode corresponds to the line where $\theta_{\rm c}$ is constant.
This is because $\theta_{\rm d}$ is uniquely determined for a given eccentricity (Eq.~\ref{eq:duty_edgeon}) and $\theta_{\rm d}=\theta_{\rm c}$ is one of the conditions for the funnel mode (Eq.~\ref{eq:funnel1}).
The left ends of these lines (the cyan line) are determined by the condition that the binary system is optically thick, $r_{\rm ph}> R_{\rm f}$, at $f=\pi\pm\theta_{\rm d}$ (Eq.~\ref{eq:funnel2}).
The right ends (part of the black line) are determined by the condition that the length of the spiral arm is longer than the photospheric radius, $r_{\rm s}>r_{\rm ph}$, at $f=\pi\pm\theta_{\rm d}$ (Eq.~\ref{eq:funnel2}).
The region allowed by the $\tau$-crossing mode corresponds to the line where $r_{\rm ph}=R_{\rm f}$ at $f=\pi\pm\theta_{\rm d}$ (Eq.~\ref{eq:tau2}).
The right ends of these lines (the cyan line) are determined by the condition that the funnel is small enough that the FRBs emitted in the inactive phase are not observable (Eq.~\ref{eq:tau1}).
This line coincides with the line where the binary system becomes optically thick in the funnel mode.
The area allowed in the $\tau$-crossing mode does not have a left edge.
This is because no matter how low $\dot{M}$ is chosen, by choosing a low $L_{\rm w}$, the funnel size becomes smaller.
Then, the FRBs are scattered by the induced Compton scattering and the periodicity can be realized.
The leftmost black line shows the parameters that realize the periodicity of FRB 121102 for the case of $e=0$.
The black line on the right of the yellow-colored region corresponds to the parameters for the case of $e=0.95$.
The light pink line is the line where $\Delta{\rm DM}=6\,{\rm cm^{-3}\,pc}$.
The right side of this pink line is excluded because $\Delta{\rm DM}$ becomes higher than the observed upper limit.
The green dashed line represents $L_{\rm w}=L_{\rm w,lower}$, and the green dash-dotted line represents $L_{\rm radio}=\dot{M}V^2/2$.
To energize the persistent radio counterpart by wind, $L_{\rm w}$ must be higher than the green dashed line (energized by the FRB pulsar wind) or $\dot{M}$ must be higher than the green dash-dotted line (energized by the companion wind).
The magenta dashed line is the line where $\beta=L_{\rm w}/\dot{M}Vc=1$ is satisfied.
Above this line, the inverse funnel mode is realized.
In summary, if the persistent radio counterpart is energized by wind, the allowed region is as follows: the colored region to the left of the light pink line and above the green dashed line for the effect of the radio efficiency, the colored region between the light pink line and the green dash-dotted line\footnote{See Sec.~\ref{sec:persistent}}, and the region above the magenta dashed line.

For the SMBH case (Fig.~\ref{fig:allowed_SMBH}), there are allowed regions.
Two different wind velocity cases, $V=3\times10^{-2}c$ and $V=1\times10^{-1}c$ are shown.
In both cases, there are allowed regions in the inverse funnel mode where the persistent radio counterpart is energized by the disk wind or the FRB pulsar wind.
In the case of $V=3\times10^{-2}c$ (left panel of Fig.~\ref{fig:allowed_SMBH}), the maximum value of $\frac{1}{2}\dot{M}V^2$ for the funnel mode or $\tau$-crossing mode is about one hundred times higher than $L_{\rm radio}$.
Thus, if the AGN activity of the SMBH is responsible for the persistent radio counterpart, the emission efficiency in the radio band must be higher than $\sim0.01$.
This value is higher than the adopted radio efficiency introduced in the last paragraph in Sec.~\ref{sec:persistent}, and so the disk wind cannot energize the persistent radio counterpart in these modes.
Thus, only the FRB pulsar wind can be responsible for the persistent radio counterpart.
In the case of $V=1\times10^{-1}c$ (right panel of Fig.~\ref{fig:allowed_SMBH}), the allowed region becomes larger than that for $V=3\times10^{-2}c$.
The AGN activity can explain the persistent radio counterpart as long as 
the emission efficiency in the radio band is higher than $\sim4\times10^{-4}$, that is, the disk wind can energize the persistent radio counterpart in these modes.
The FRB pulsar wind can also energize the persistent radio counterpart.
For the SMBH case, if the radio emission originates from 
the jet, the green dash-dotted line can move to the left and the allowed region becomes larger.
The stellar mass of the host galaxy of FRB 121102 is about $(4$--$7)\times10^{7}M_\odot$.
The spectra of dwarf galaxies with such small masses are not well understood.
Thus, 
the SMBH at a slight offset from the galaxy center 
could be responsible for 
the persistent radio counterpart and its observed spectrum.

For the MS case (Fig.~\ref{fig:allowed_MS}), two different wind velocity cases, $V=1\times10^{-2}c$ (the observed maximum velocity) and $V=3\times10^{-2}c$ (extremely high velocity) are shown.
In both cases, there are allowed regions in the inverse funnel mode where the persistent radio counterpart would be energized by the FRB pulsar wind.
An extremely high mass loss rate ($\gtrsim10^{-1.5}M_\odot\,{\rm yr^{-1}}$) is required to energize the persistent radio counterpart by the companion wind with a radio efficiency of $10^{-3}$.
If $V=3\times10^{-2}c$, there are allowed region in the funnel mode where the persistent radio counterpart is energized by the FRB pulsar wind (right panel of Fig.~\ref{fig:allowed_MS}).
If $V=1\times10^{-2}c$, in the allowed region for the funnel mode or $\tau$-crossing mode, the FRB pulsar wind luminosity is too weak to reproduce the persistent radio counterpart (left panel of Fig.~\ref{fig:allowed_MS}).
Furthermore, the kinetic luminosity of the companion wind is lower than the radio luminosity of the persistent radio counterpart.
Thus, the MS case with typical velocities $V \lesssim 1\times 10^{-2} c$ is excluded in the funnel mode and $\tau$-crossing mode.

For the IMBH case (Fig.~\ref{fig:allowed_IMBH}), there are allowed regions in the cases of $V=1\times10^{-1}c$ and $V=3\times10^{-2}c$.
In both the velocity cases, the inverse funnel mode is possible and the persistent radio counterpart is energized by the activity of the IMBH or the FRB pulsar wind.
In the case of $V=1\times10^{-1}c$, the persistent radio counterpart is energized by the synchrotron nebula or by the activity of the IMBH in the funnel mode or $\tau$-crossing mode.
In the case of $V=3\times10^{-2}c$, the disk wind cannot be responsible for the persistent radio counterpart in the funnel mode or $\tau$-crossing mode if the radio efficiency is taken into account, while the FRB pulsar is able to energize the persistent radio counterpart.
As these figures show, the allowed region is highly dependent on the wind velocity of the companion wind.

The beaming effect of each FRB loosens the constraints on the parameter space.
In the funnel mode, if FRBs are beamed, the half-opening angle of the funnel, $\theta_{\rm c}$, must be larger than $\theta_{\rm d}$.
This is because the beaming effect makes the active window shorter than that without the beaming effect.
For a given eccentricity, on the line of $\theta_{\rm d}=\theta_{\rm c}$ (constant eccentricity line for the funnel mode; see Figs.~\ref{fig:allowed_SMBH}--\ref{fig:allowed_IMBH}), the observed duty cycle is realized without the beaming effect.
For that eccentricity, between this line and the magenta dashed line, we have $\theta_{\rm d}<\theta_{\rm c}$ and the active window is larger than that on the $\theta_{\rm d}=\theta_{\rm c}$ line.
Therefore, for a given eccentricity, the region between the line of $\theta_{\rm d}=\theta_{\rm c}$ and the magenta dashed line is also allowed if the beaming effect is taken into account.
In the $\tau$-crossing mode, if FRBs are beamed, $\theta_{\rm d}$ must be larger than the value for the unbeamed case.
Thus, for a given eccentricity, the allowed region is to the left of the line allowed in the unbeamed case.
The region to the right of the colored region is not allowed even if the beaming effect is taken into account.

If we consider the general viewing angle, there are two effects on the allowed region for the funnel mode.
One effect is that the allowed region gets smaller.
First, we consider the effect of changing the argument of periapsis.
For given $\theta_{\rm c}$ (i.e., $L_{\rm w}$ and $\dot{M}$) and $e$, if the argument of periapsis is changed, the duty cycle gets shorter (see Figs.~\ref{fig:e_d_025} and \ref{fig:e_d_01_05} in Appendix~\ref{sec:general_angle}).
To make the duty cycle longer, the eccentricity must get higher.
If there is an eccentricity which realizes the observed duty cycle, these parameters are also allowed.
However, if there is no such eccentricity, these parameters are not allowed in the given argument of periapsis.
Next, changing the inclination angle also makes the required eccentricity higher because the half-opening angle of the funnel effectively becomes smaller (see Appendix~\ref{sec:general_angle}).
The other effect is that the disallowed region between the magenta dashed line and the colored region in Figs.~\ref{fig:allowed_SMBH}--\ref{fig:allowed_IMBH} might become an allowed region.
This region is not allowed because the half-opening angle is too large and the duty cycle is larger than the observed duty cycle.
For given $\theta_{\rm c}$ and $e$, if the argument of periapsis is changed, the duty cycle gets shorter (see Figs.~\ref{fig:e_d_025} and \ref{fig:e_d_01_05} in Appendix~\ref{sec:general_angle}).
Thus, the observed duty cycle might be realized and these parameters becomes allowed.
Changing the inclination angle also has the same effect (see Fig.~\ref{fig:e_d_01_05} in Appendix~\ref{sec:general_angle}).
Even in this case, the region to the right of the colored region is not allowed.

For the $\tau$-crossing mode, the allowed region would get smaller if we consider the general viewing angle.
For a given $e$, if the argument of periapsis is changed, the region near the apoapsis gets optically thicker.
Because the FRB pulsar spends more time near the apoapsis, the active window gets shorter.
As a result, the allowed region would get smaller from the same discussion as in the funnel mode.
Changing the inclination angle has the same effect.
This is because there are more wind particles along the line of sight to the observers in $I\neq \pi/2$.

\section{Frequency-dependence of the active window in the binary comb model}\label{sec:frequency}
Recent observations suggest that the beginning and the width of the active window of FRB 180916 depend on the 
observed frequency.
\cite{Ple2020} compared the observations by LOFAR high-band antenna (110--190 MHz), the uGMT (200--450 MHz), and the CHIME (400--800 MHz).
\cite{Pas2020} compared the observations by the LOFAR, the CHIME, and the Apertif (1220--1520 MHz) 
and reported that the peak activity and the width of the active phase vary with the frequency (their Fig.~4).
In their observations, the active window in the Apertif band begins at the same time as that in the CHIME band but ends earlier than in the CHIME band.
The peak activity in the Apertif band is $\sim$0.7 days before that in the CHIME band.
The full-width at half-maximum is 1.1 days in the Apertif band while it is 2.7 days in the CHIME band.
The active phase in the LOFAR band begins later than that in the CHIME band and ends later than in the CHIME band.
The peak activity in the LOFAR band is $\sim$2 days after that of the CHIME, and the activity window is not estimated because of its lower number of samples.

The observed earlier active phase in a higher frequency band is actually predicted in the binary comb model.
This is because higher-frequency FRBs are less obscured by the plasma absorption or scattering in the companion wind.
On the other hand, the observed
earlier ending of the active window in a higher frequency band does not seem to support the binary comb model and \cite{Ple2020,Pas2020} have used this effect to disfavor the binary comb model and support a magnetar precession model. 
In this section, we revisit this problem and argue that it is premature to disfavor the binary comb model using the observational results of \cite{Ple2020} and \cite{Pas2020}. We suggest two scenarios in which a shorter active window is realized in the higher frequency band.

\begin{figure}[H]
  \begin{center}
    \begin{tabular}{c}
      \begin{minipage}[t]{0.5\hsize}
        \begin{center}
          \includegraphics[clip,width=\textwidth]{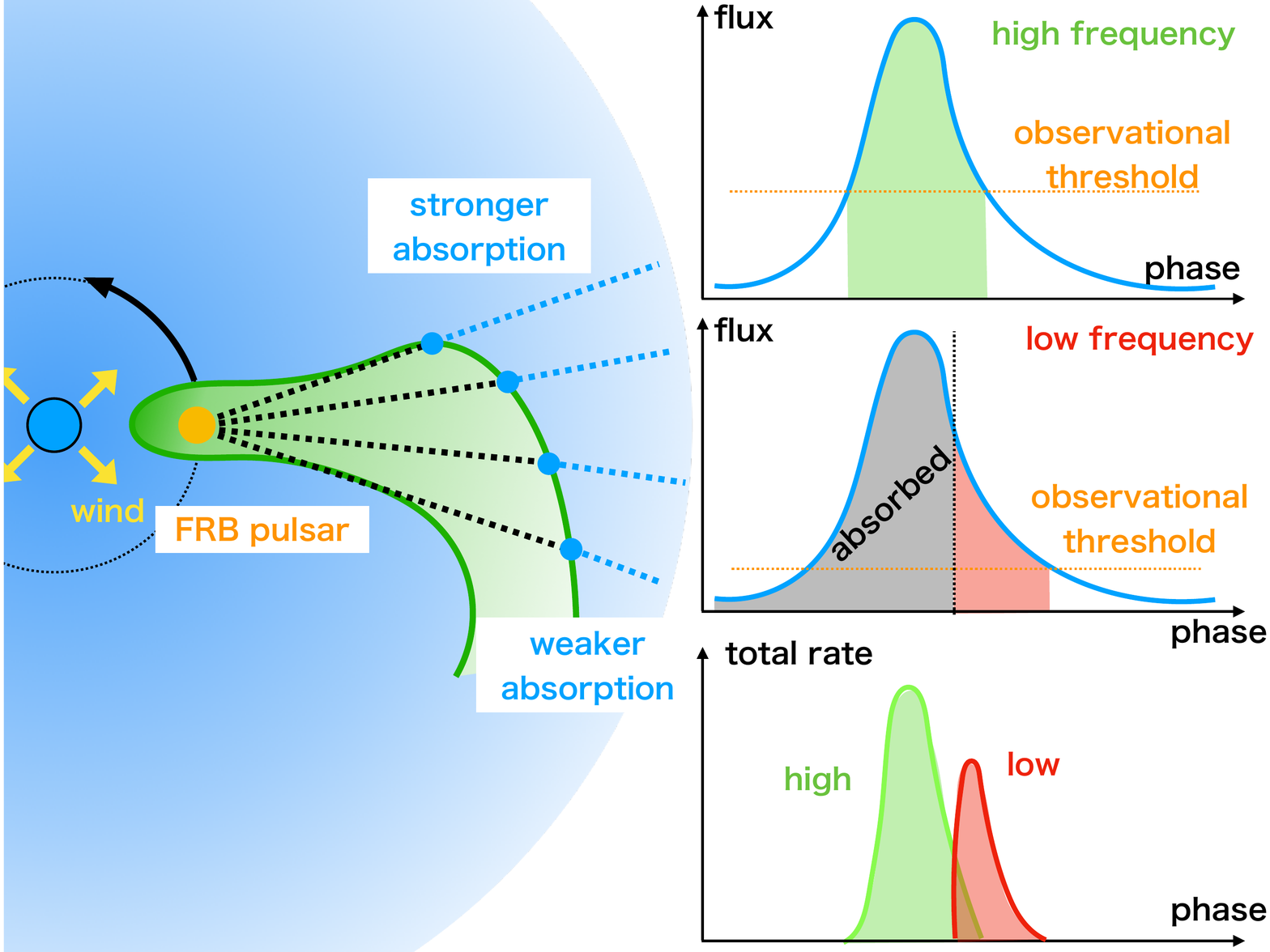}
          \caption{
          The observational threshold and funnel configuration that realizes the observed frequency dependent active window of FRB 180916 (the sensitivity-absorption scenario).
          The dotted lines in the left side show the line of sight to the observer at different phases.
          The blue line in the right panel shows the intrinsic flux distribution profile as a function of phase.
          At an earlier phase, the FRBs are exposed to a denser wind so that low-frequency bursts (the gray region in the right middle panel) are absorbed.
          The yellow dotted lines in the right panel show the observational thresholds for different bands and FRBs below these lines are not observable.
          Both the observational threshold and absorption contribute to create the observed frequency-dependent active window.
          }
          \label{fig:sensitivity}
        \end{center}
      \end{minipage}
      
      \begin{minipage}[t]{0.5\hsize}
        \begin{center}
          \includegraphics[width=\textwidth]{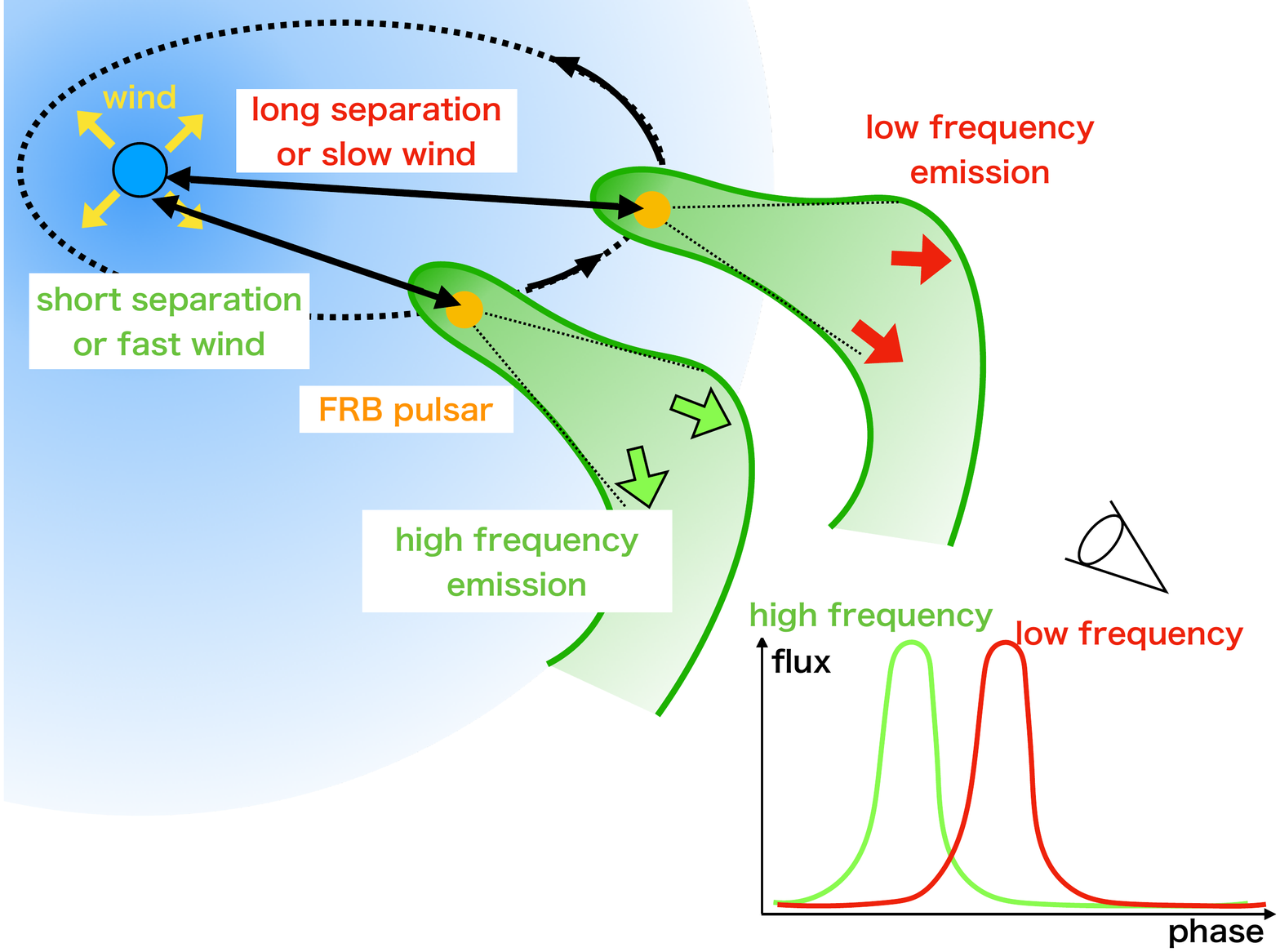}
          \caption{
          The orbital separation dependence or the wind velocity dependence of the FRB frequency that make the observed frequency dependent active window of FRB 180916 (the aurora scenario).
          When the separation is smaller, more aurora particles would reach the inner part of the magnetosphere and their number density is higher.
          The plasma frequency is also higher and the frequency of the FRBs is also higher if they are emitted around the plasma frequency.
          Alternatively, when the wind velocity is higher, the aurora particles reach more inner part of the magnetosphere.
          If the curvature radiation occurs there, the frequency of the FRBs would be higher because of the larger curvature.
          One of these scenarios can reproduce the observed frequency-dependent active window.
          }
          \label{fig:separation}          
        \end{center}
      \end{minipage}
    \end{tabular}
  \end{center}
\end{figure}

The first possibility is the sensitivity-absorption scenario.
The earlier ending of the active window in the higher frequency band is due to 
the different sensitivities of different telescopes, with the threshold sensitivity in the higher frequency band shallower than in the lower frequency band.
Let the intrinsic flux distribution profile has some structure in phase as shown in the blue line in the right part of Fig.~\ref{fig:sensitivity}.
The sensitivities of the telescopes normalized by the flux profile are different between the different telescopes (the right top of Fig.~\ref{fig:sensitivity}).
Only FRBs whose fluxes are higher than the observational threshold are
observable.
If the sensitivity of the telescope in the low frequency is higher than that in the high frequency, 
fainter FRBs are observable in low frequency not in high frequency (the right middle of Fig.~\ref{fig:sensitivity}).
In addition to this observational threshold, low-frequency FRBs radiated in the earlier orbital phases are blocked by the wind because they propagate in a denser wind (the left side of Fig.~\ref{fig:sensitivity}).
Because the event rate is generally higher for fainter events, the final detected event rate in each frequency 
becomes frequency dependent as observed, as shown
in the right bottom of Fig.~\ref{fig:sensitivity}.
This scenario is supported by the fact that the longest active phase is in the CHIME band where the sensitivity of the telescope is high, and by the fact that the phase of the active window is not monotonically dependent on frequency.
In addition, we should note that the active window for each frequency also depends on the spectrum of the burst.

The second possibility is the aurora scenario.
In this scenario, the shorter active window in the higher frequency is due to the intrinsic emission mechanism (Fig.~\ref{fig:separation}).
As discussed in \cite{IokZha2020}, the binary interaction might provide the condition to facilitate the FRB coherent radiation mechanism.
Although the contact discontinuity of the funnel is larger than the magnetosphere of the FRB pulsar, some of the wind particles may enter the magnetosphere through the interface of the contact discontinuity of the funnel
and the light cylinder of the FRB pulsar, forming an aurora-like inflow similar to Earth. 

The change in the number of the aurora particles may change the FRB frequency.
One of the factors that may change the FRB frequency is the eccentricity.
If the binary of FRB 180916 has non-zero eccentricity, the separation of the binary changes with time.
When the separation is small, the number density of the companion wind around the contact discontinuity is high.
When the number density is high, the number of the aurora particles would be large.
Therefore, the smaller the separation, the more aurora particles are expected to be present.
When the amount of the aurora particles is large, the plasma frequency around the emission area would be high ($\nu_p\propto n_{\rm a}^{1/2}$, where $\nu_{\rm p}$ is the plasma frequency, $n_{\rm a}$ is the aurora particle number density).
If FRBs are emitted due to plasma effects, the FRBs are expected to be emitted around the plasma frequency.
Therefore, when the separation is small, the FRB frequency would be high.
Because the separation changes with time, the FRB frequency also changes with time.

The FRB frequency may also change as the velocity of the companion wind received by the FRB pulsar changes with time.
The velocity of the companion wind can vary with latitude, as in the case of the Sun.
Thus, if the spin of the companion star and the orbital angular momentum of the binary are not parallel, the velocity of the companion wind received by the FRB pulsar can vary with time.
The higher the velocity of the companion wind, the higher the kinetic energy of the wind particles.
This high kinetic energy allows the aurora particles to reach the more inner regions of the magnetosphere.
The reason is as follows.
In the inner region of the magnetosphere, the magnetic field is stronger, and the energy at the lowest Landau level becomes higher.
Because the magnetic field does not perform any work on the aurora particles, the total energy of the particles does not increase.\footnote{The total energy is rather reduced by synchrotron radiation.}
Therefore, the particles cannot penetrate inside the region where the total energy of the particles is equal to the energy of the lowest Landau level.
Thus, the radius where the aurora particles can reach, $R_{\rm in}$, is given by solving the  equation\footnote{We do not take  synchrotron radiation and the magnetic mirror effect into account. 
With these effects, $R_{\rm in }$ becomes larger than this value.}
\be
\frac{1}{2}m_{\rm e} V^2=\frac{1}{2}\hbar\frac{eB(R_{\rm in})}{m_{\rm e}c},
\ee
where $B(R_{\rm in})$ is the magnetic field at $R_{\rm in}$.
Assuming a dipolar magnetic field, we obtain
\be
R_{\rm in }\simeq1\times10^{7}\left(\frac{V}{0.1c}\right)^{-2/3}\left(\frac{B_{\rm NS}}{10^{15}\, {\rm G}}\right)^{1/3}\left(\frac{R_{\rm NS}}{10^6\,{\rm cm}}\right)\,{\rm cm},
\ee
where $B_{\rm NS}$ is the magnetic field at the neutron-star surface and $R_{\rm NS}$ is the neutron-star radius.
If the aurora particles trigger the FRB radiation via curvature radiation, the FRB frequency ($\propto \gamma^3/\rho_{\rm c}$, where $\gamma$ is the Lorentz factor of the particle and $\rho_{\rm c}$ is the magnetic field curvature radius) becomes higher in the more inner region of the magnetosphere
\citep{LiZan2021}.

\section{Summary and discussion}\label{sec:summary}
In this paper, we develop the binary comb model to constrain the host binary for FRB 121102.
We found that for each type of companion (MS, IMBH, SMBH), there are some parameters in which the observed periodicity, change in the dispersion measure, and the persistent radio counterpart of FRB 121102 are realized.
An SMBH with disk wind is one of the best candidates for the companion of FRB 121102.
An IMBH is also a possible candidate.
An MS is found to be an unfavorable companion because the stellar wind is too weak to power the persistent radio counterpart of FRB 121102.
On the other hand, if the persistent radio counterpart is energized by the FRB pulsar, the MS companion is still viable because the binary may be in the inverse funnel mode (see Fig.~\ref{fig:allowed_MS}).

We have extended the binary comb model in two ways: by taking into account the eccentricity of binary, and by considering 
the other modes not considered in the original model of \cite{IokZha2020}.
The addition of eccentricity to the model plays an essential role in explaining periodic FRBs with a large active window.
In addition to the funnel mode (see Fig.~\ref{fig:funnel}a) that has been considered in \cite{IokZha2020}, we show that there are two other modes.
One is the $\tau$-crossing mode in which the period is explained by the temporal variation of the optical depth (see Fig.~\ref{fig:funnel}b).
The other is the inverse funnel mode in which the funnel is reversed (see Fig.~\ref{fig:funnel}c).

For given companion mass and velocity of the companion wind, the allowed parameter space in the $\dot{M}$-$L_{\rm w}$ plane is constrained by the observed period, duty cycle, the change in the dispersion measure, and the ability to provide enough energy for the persistent radio counterpart.
For any type of companion, there are parameter regions where the observed signatures are realized in the inverse funnel mode.
In the parameters where the periodicity and the duty cycle are realized in the funnel mode or $\tau$-crossing mode, the other constraints (change in the dispersion measure and persistent radio counterpart) are important.
In the MS companion case with a typical velocity $V=1\times10^{-2}c$, there are some parameters where the change in the dispersion measure is less than the observed value in the funnel mode and $\tau$-crossing mode.
However, the persistent radio counterpart is not as bright as observed in these parameters.
Therefore, these modes are excluded in this case, 
although they are allowed
in the extremely high-velocity case with $V=3\times10^{-2}c$.
In the SMBH or IMBH companion case with $V=1.0\times10^{-1} c$ (see also \cite{Zha2018}), there is a parameter region in the funnel mode and $\tau$-crossing mode where the upper limit on the change in the dispersion measure is satisfied and the persistent radio counterpart is as bright as observed. 
On the other hand, if the radio emission originates from
the jet of the central black hole, the cases of $V=3.0\times10^{-2} c$ are not excluded.
It was found that a high velocity of the wind is important to be the companion of FRB 121102.

We have also proposed two scenarios to explain the frequency dependence of the active window of FRB 180916.
In the sensitivity-absorption scenario, the earlier ending of the active window in a higher frequency is due to the different sensitivities of different telescopes, and the later onset of the active phase in a low frequency is due to  absorption.
In the aurora scenario, the frequency-dependent active window is explained by the change in the amount of the aurora particles in the magnetosphere.
The separation of the binary or the wind velocity of the companion can change in orbital phases, and this change invokes a variation in the amount of aurora particles.
Elaboration of these models is future work.

In the SMBH or IMBH companion case, the accretion disk of the black hole can affect the periodicity of the FRBs.
The size of the accretion disk is $\sim10^3$--$10^4R_{\rm Sch}$ for the AGN disk where $R_{\rm Sch}$ is the Schwarzschild radius of the black hole \citep{Net2015,LaoNet1989}.
This could be of the same order as the semi-major axis but is smaller than the semi-major axis.
Therefore, it would not affect the active window in the case of FRB 121102.
If the disk is larger than the shortest separation in the active window and $\theta_{\rm d}$ is larger than $\pi/2$, the accretion disk can affect the periodicity.
Also, the accretion disk has the potential to shine on its own.
The upper limit on the luminosity of the persistent counterpart in optical band and in X-ray band is $\sim10^{41}\,{\rm erg\,s^{-1}}$ \citep{Cha2017}.
This value is $\sim10^{-2}L_{\rm Edd}$ for the SMBH case and $\sim L_{\rm Edd}$ for the IMBH case.
These are the upper limits on the luminosity of the black hole accretion disk.

In the future, it is expected that more periodic FRBs may be detected and they can help to determine the origin of the periodicity, e.g. binary or precession.
In the binary comb model, the change in the dispersion measure $\Delta {\rm DM}$ may be also periodic,
providing a possible smoking-gun signature for this model. 
Furthermore, the variation of the rotation measure may be also periodic as proposed in \cite{Zha2018}.
Because the contribution to the rotation measure would be larger near the source than in the intergalactic medium, the periodic change in the rotation measure may be more prominent and likely be observable. Long-term RM variation of FRB 121102 has been observed showing a complicated behavior \citep{Hil2021}, and the binary scenario has not been applied to compare against the data. 
In the binary comb model, $\dot{M}$ and $V$ are important in determining the active window, and if they change during the observation, the periodicity may also change accordingly.
In this case, the active window can change while the period does not change.
Also, in the future, we might observe a periodic FRB from a binary in three body systems.
In this case, the period would be both increasing and decreasing in the binary comb model.
On the other hand, in the precession model, the period is either increasing or decreasing monotonically \citep{Kat2021_periodic}.
If future periodic FRBs exhibit both increasing and decreasing periods, the origin of the period might be related to binaries in the three body systems.

In the SMBH or IMBH companion case, the persistent radio counterpart is expected to be created by the outflow from the black hole.
It is an open question whether the radio emission due to the outflow can reproduce the observed spectrum.
In the sensitivity-absorption scenario, the intrinsic flux distribution profile is assumed and its origin needs to be clarified.
In the aurora scenario, the relationship between the FRB frequency and the separation or between the FRB frequency and the wind velocity needs to be clarified.
These open questions will be studied in future works.

\section*{ACKNOWLEDGMENTS}
We are grateful to Kazuya Takahashi, Hamidani Hamid, Wataru Ishizaki, Koutarou Kyutoku, Susumu Inoue, Kazumi Kashiyama, Norita Kawanaka, Kohta Murase, and Shuta Tanaka, for fruitful discussion and valuable comments.
We also thank the participants and the organizers of the workshops with the identification number YITP-T-20-04 for their generous support and helpful comments.
This work is supported by Grants-in-Aid for Scientific Research No. 20J13806 (TW), 20H01901, 20H01904, 20H00158, 18H01213, 18H01215, 17H06357, 17H06362, and 17H06131 (KI) from the Ministry of Education, Culture, Sports, Science and Technology (MEXT) of Japan.

\appendix
\section{General viewing-angle case }\label{sec:general_angle}
\begin{figure}[H]
\begin{center}
          \includegraphics[clip, width=0.6\textwidth]{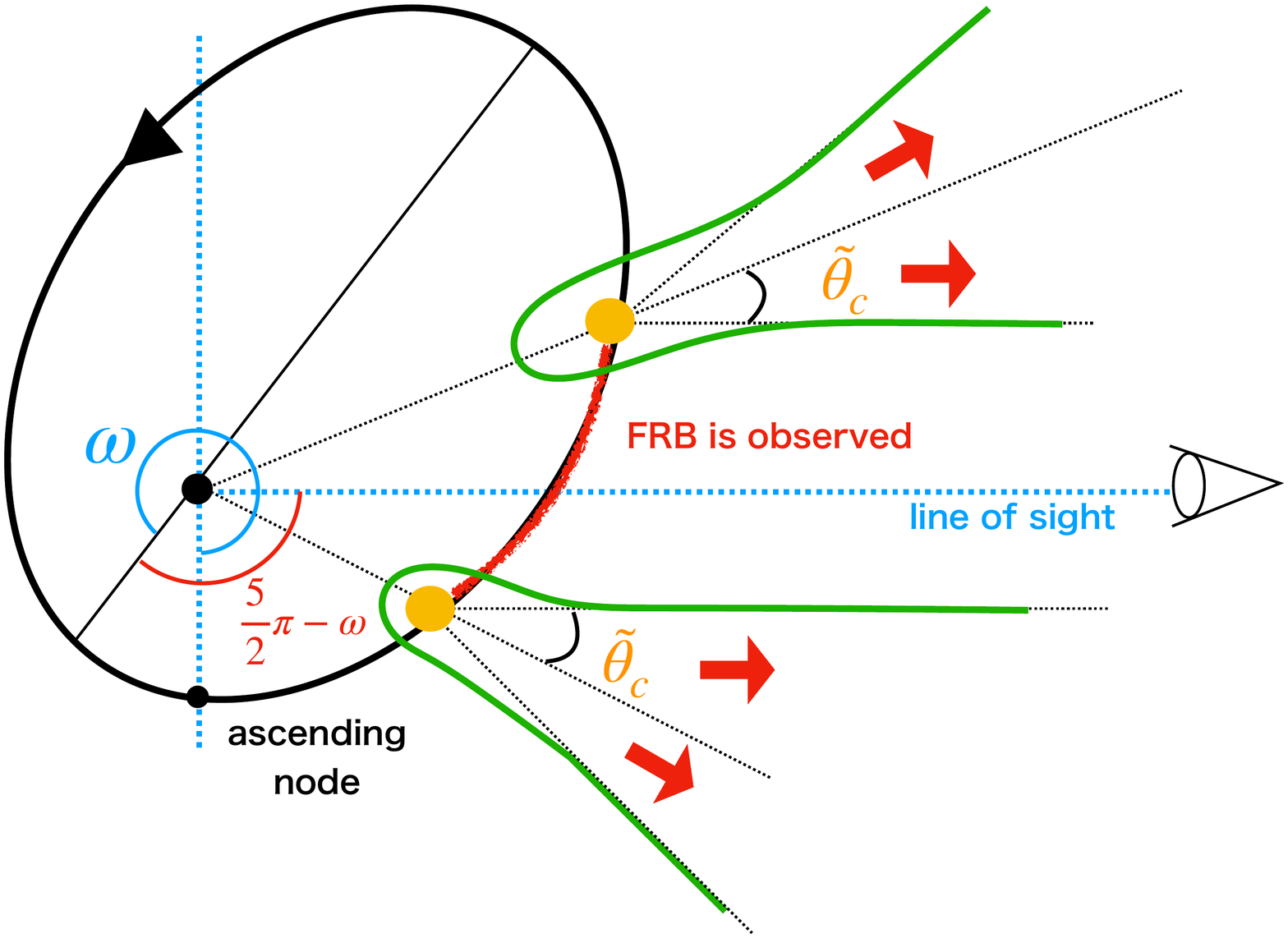}
              \caption{
              An overview of the funnel mode in a general viewing angle. 
              The black ellipse with the black arrow is the orbit of the FRB pulsar, and the black circle in the ellipse is the center of mass.
              $\omega$ is the argument of periapsis which is the angle between the ascending node and the periapsis.
              The green lines represent the funnels and the FRBs through the funnel are observable.
              When the FRB pulsar is on the red curve in the ellipse, the FRBs are observable.
              In this figure, the line of sight is projected onto the orbital plane.
              }
              \label{fig:ov}
  \end{center}
\end{figure}

In this Appendix, we consider the duty cycle for an arbitrary viewing angle (Fig.~\ref{fig:ov}) for the funnel mode.
Generally, the configuration of a binary is determined by an orbital inclination angle, $I$, and an argument of periapsis, $\omega$.
The fundamental plane of the binary is orthogonal to the line of sight and the inclination angle, $I$, is defined as the angle between the fundamental plane and the orbital plane.
The argument of periapsis $\omega$ is the angle from the ascending node to the periapsis, and the ascending node is where the star ascends from the fundamental plane.
In the main text, we have adopted $I=\pi/2$ and $\omega=3\pi/2$.

\begin{figure}[H]
\begin{center}
          \includegraphics[clip, width=0.6\textwidth]{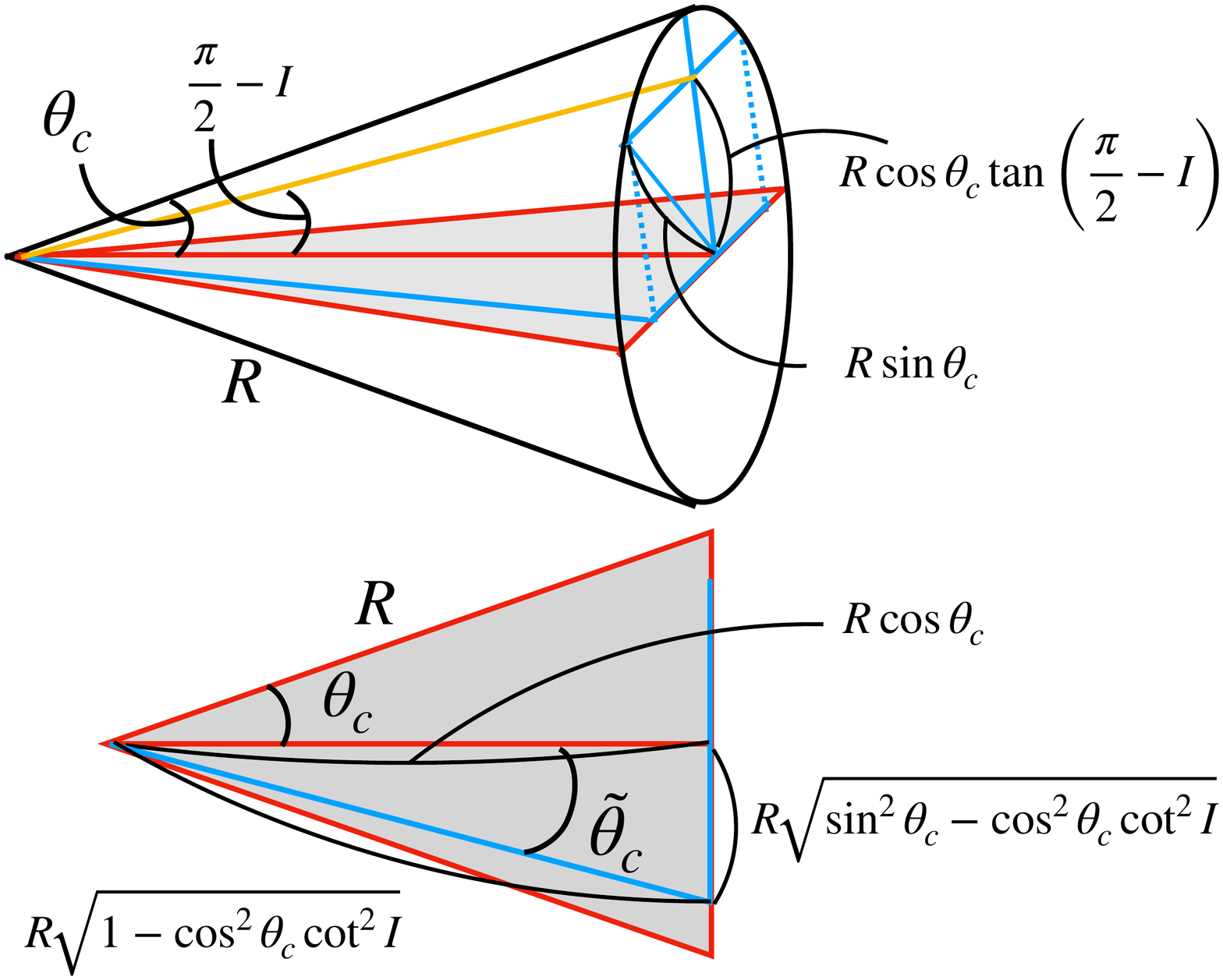}
              \caption{
              The effective half-opening angle of the funnel, $\tilde{\theta_{\rm c}}$.
              The shape of the funnel is assumed to be a cone (top panel).
              The shaded triangle is on the orbital plane, and the yellow line represents the line of sight.
              The triangle in the bottom panel is the shaded triangle in the top panel.
              For given $\omega$ and $I$, $\tilde{\theta_{\rm c}}$ is expressed by $\theta_{\rm c}$.
              }
              \label{fig:theta_c}
  \end{center}
\end{figure}

First, we evaluate the effect of changing the orbital inclination angle, $I$.
For $I\neq\pi/2$, the half-opening angle of the funnel becomes effectively smaller than that for $I=\pi/2$ (Fig.~\ref{fig:theta_c}).
Thus, we introduce an effective half-opening angle, $\tilde{\theta_{\rm c}}$, as shown in Fig.~\ref{fig:theta_c}.
We assume that the shape of the funnel is a cone whose apex angle equals $\theta_{\rm c}$.
Then, $\tilde{\theta_{\rm c}}$ is related to the orbital inclination angle, $I$, and the half-opening angle of the funnel, $\theta_{\rm c}$, through the equation
\be
\cos\tilde{\theta_{\rm c}}=\frac{\cos\theta_{\rm c}}{\sqrt{1-\cos^2\theta_{\rm c}\cot^2 I}}.
\label{eq:theta_c_eff}
\ee
In the case of $I=\pi/2-\theta_{\rm c}$, the effective half-opening angle equals zero because the line of sight coincides with the baseline of the cone.

Next, we evaluate the effect of changing the argument of periapsis, $\omega$.
We can observe FRBs when the true anomaly is between $\frac{5\pi}{2}-\omega-\tilde{\theta_{\rm c}}$ and $\frac{5\pi}{2}-\omega+\tilde{\theta_{\rm c}}$ (see Fig.~\ref{fig:ov}).
Thus, the duty cycle is represented as 
\bea
\frac{T}{P}=\frac{1}{2\pi}\int^{u_+}_{u_-}\frac{dt}{du}du= \frac{1}{2\pi}\lsb\lrb u_+-e\sin u_+\rrb-\lrb u_--e\sin u_-\rrb+2\pi\Theta(-u_+)\Theta(u_-)\rsb,
\label{eq:duty}\\
u_{\pm}=2\tan^{-1}\lrb\sqrt{\frac{1-e}{1+e}}\tan\frac{\frac{5\pi}{2}-\omega\pm\tilde{\theta_{\rm c}}}{2}\rrb\quad (-\pi<u_\pm<\pi),
\eea
where $\Theta(x)$ is the step function. 
We note that the relation between the true anomaly and the eccentric anomaly is valid in $-\pi<f<\pi$.  
Thus, in the case of $u_->0$ and $u_+<0$, we have to calculate the integration in Eq.~(\ref{eq:duty}) separating $u_-<u<\pi$ and $-\pi<u<u_+$ and an additional $2\pi$ factor appears. 

\begin{figure}[H]
  \begin{center}
    \begin{tabular}{c}
      \begin{minipage}{0.5\hsize}
        \begin{center}
          \includegraphics[clip,width=\textwidth]{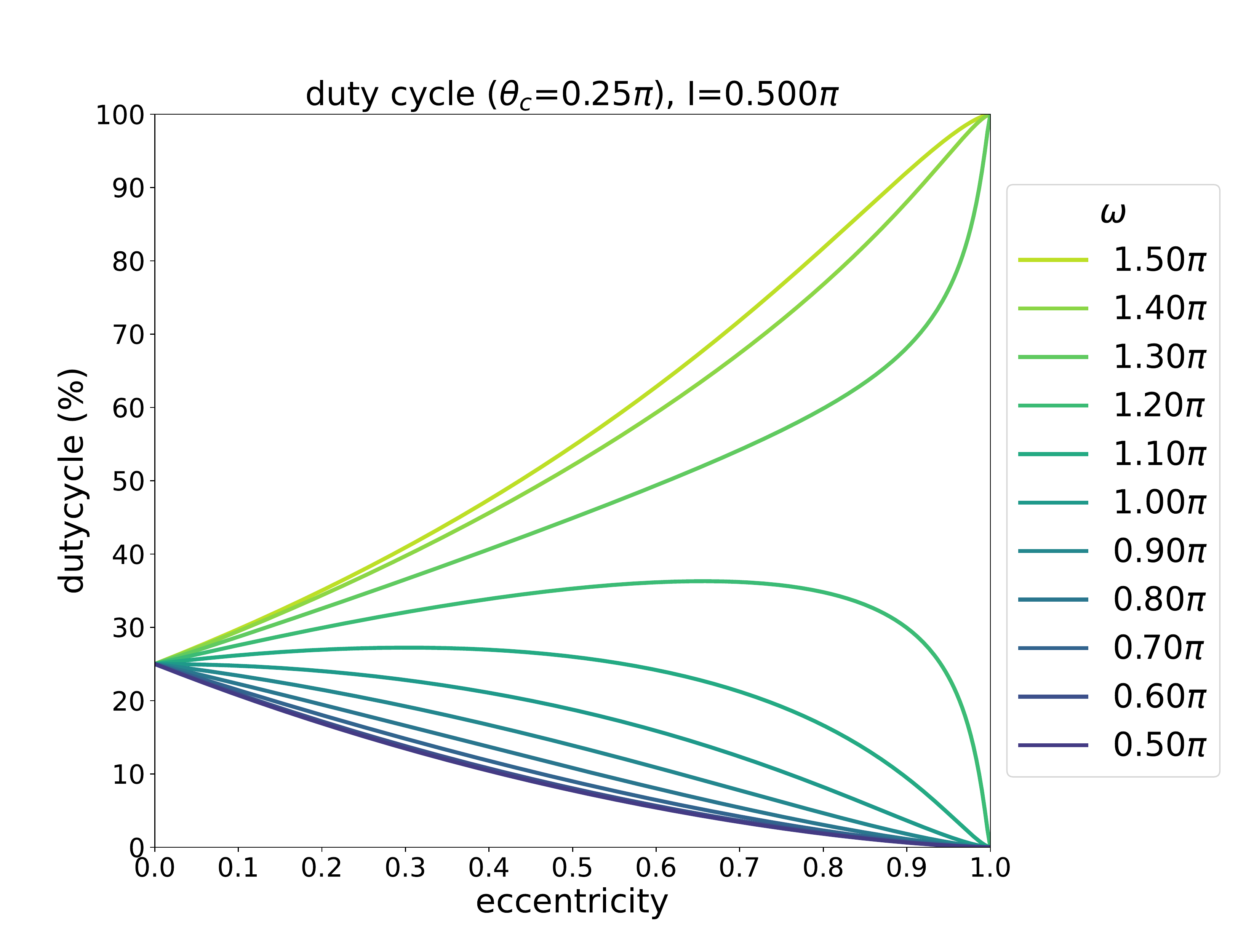}
        \end{center}
      \end{minipage}
      \begin{minipage}{0.5\hsize}
        \begin{center}
          \includegraphics[width=\textwidth]{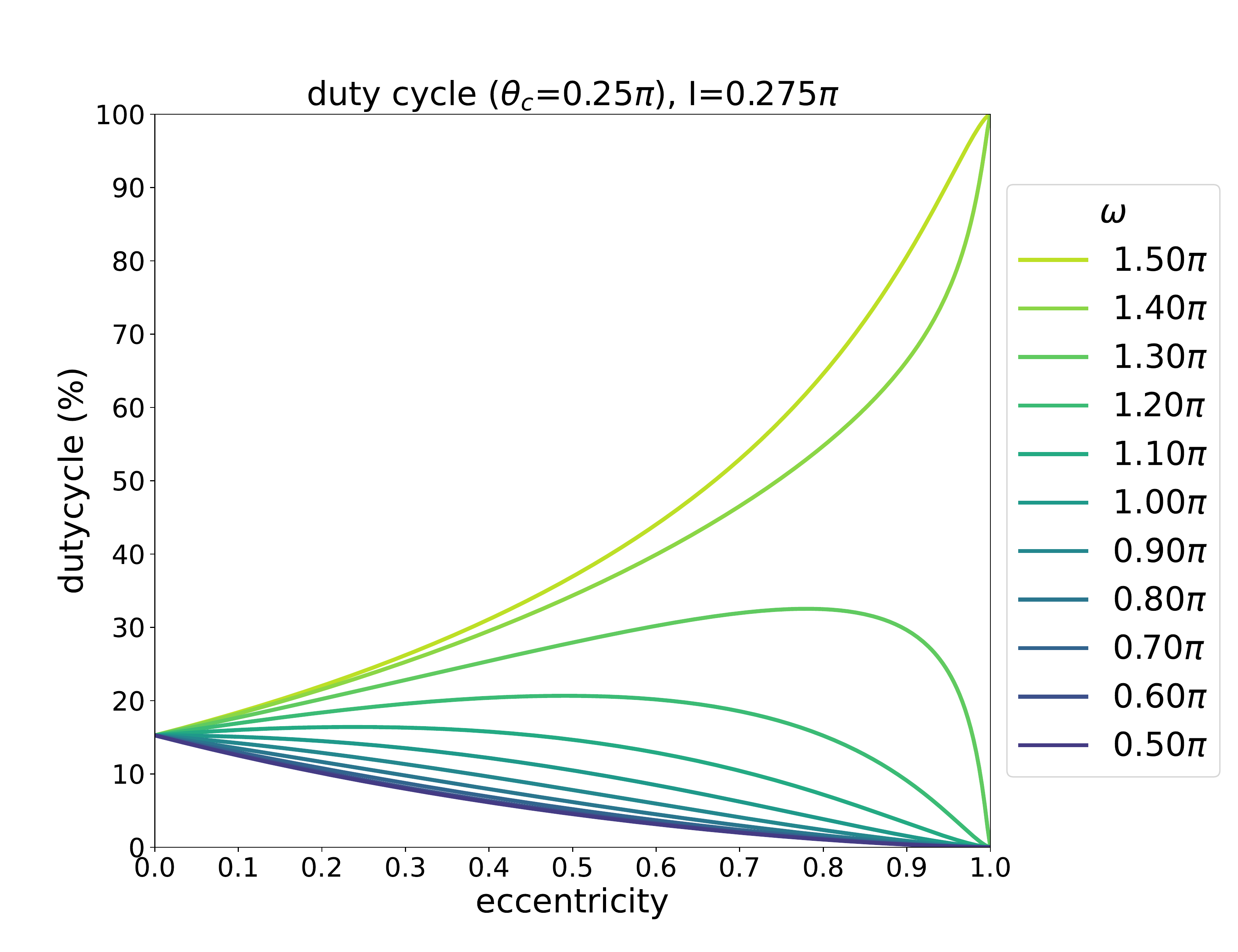}
        \end{center}
      \end{minipage}
    \end{tabular}
    \caption{
      The relations between duty cycle and eccentricity.
    The colors show the argument of periapsis, $\omega$.
    The left panel shows the case that the half-opening angle of the funnel, $\theta_{\rm c}$, equals $0.25\pi$ and the inclination angle, $I$, equals $0.5\pi$.
    The right panel shows the case that $\theta_{\rm c}=0.25\pi$ and $I=0.275\pi=0.5\pi-0.9\theta_{\rm c}$.
    }
    \label{fig:e_d_025}
  \end{center}
\end{figure}

\begin{figure}[H]
  \begin{center}
    \begin{tabular}{c}
      \begin{minipage}{0.5\hsize}
        \begin{center}
          \includegraphics[clip,width=\textwidth]{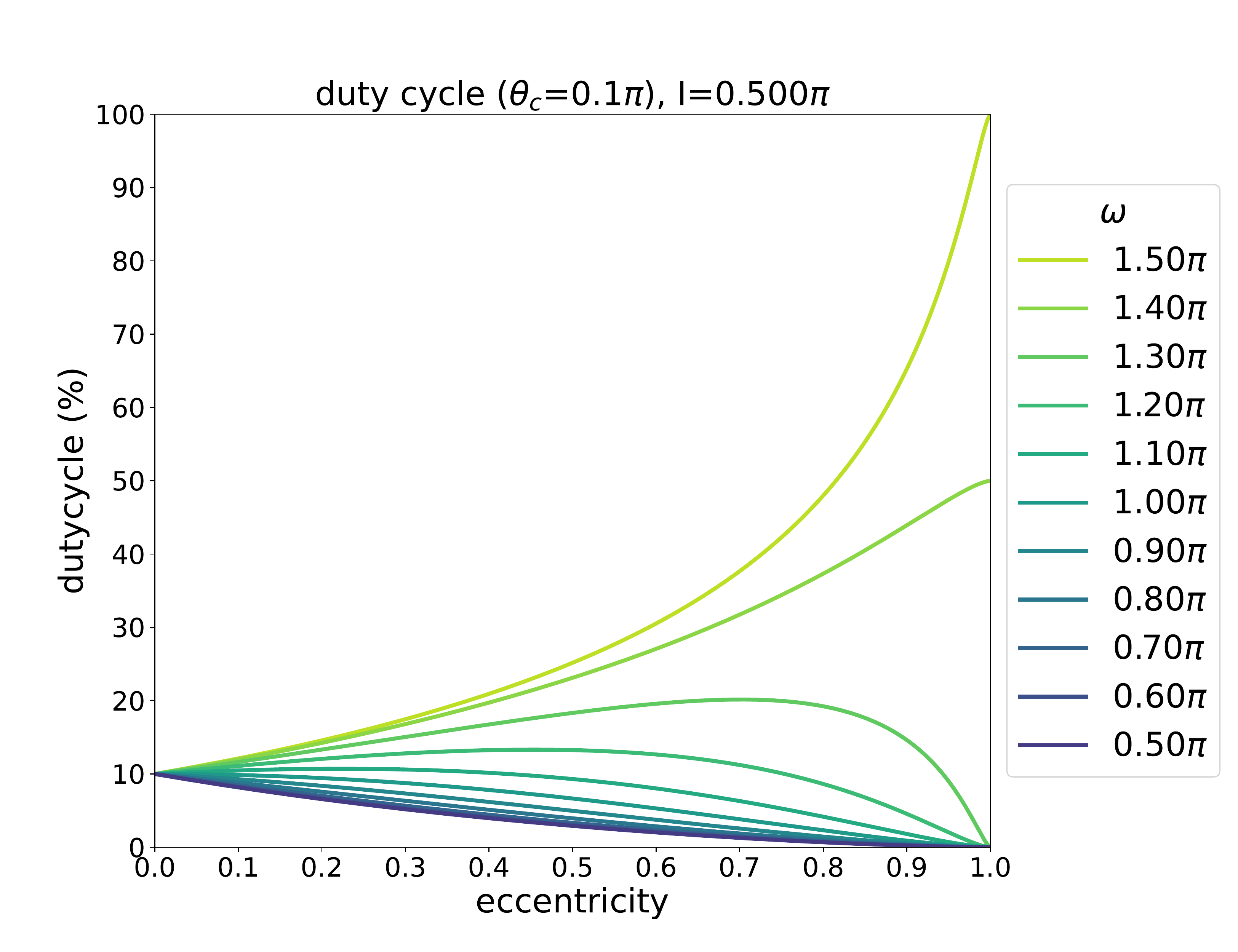}
        \end{center}
      \end{minipage}
      \begin{minipage}{0.5\hsize}
        \begin{center}
          \includegraphics[width=\textwidth]{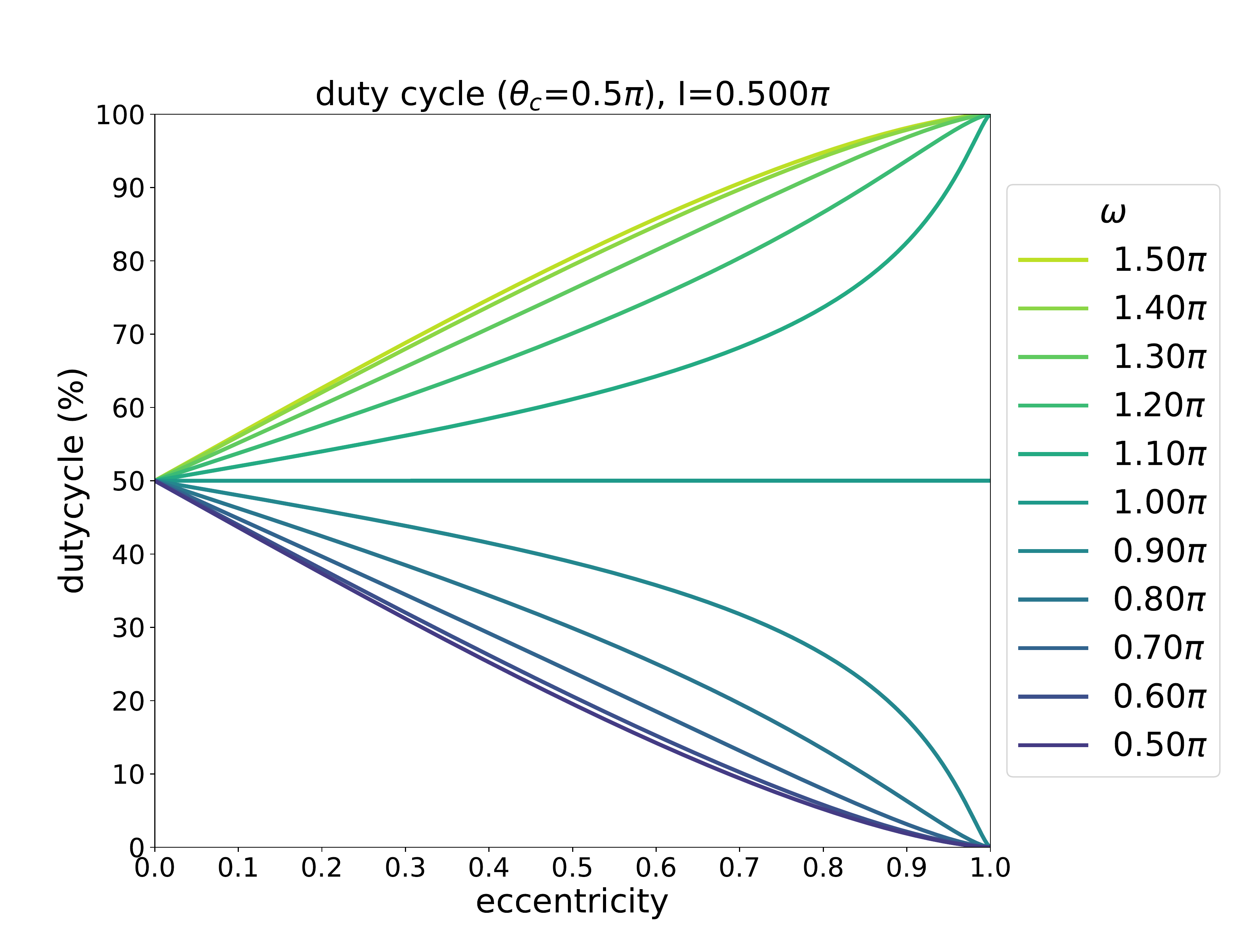}
        \end{center}
      \end{minipage}
    \end{tabular}
    \caption{
      The relations between duty cycle and  eccentricity. 
      The left panel shows the case that the half-opening angle of the funnel, $\theta_{\rm c}$, equals $0.1\pi$ and the inclination angle, $I$, equals $0.5\pi$.
      The right shows the case that $\theta_{\rm c}=0.5\pi$ and $I=0.5\pi$.
    }
    \label{fig:e_d_01_05}
  \end{center}
\end{figure}

Using Eqs.~(\ref{eq:theta_c_eff}) and (\ref{eq:duty}), we obtain the relation between the eccentricity and the duty cycle (Figs.~\ref{fig:e_d_025} and \ref{fig:e_d_01_05}).  
In each figure, we fix $\theta_{\rm c}$ and $I$, and change $\omega$ from $\pi/2$ (observation from the direction of the periapsis) to $3\pi/2$ (observation from the direction of the apoapsis).
Figure~\ref{fig:e_d_025} shows the relations for $(\theta_{\rm c},\,I)=(0.25\pi, 0.5\pi),\,(0.25\pi,0.275\pi)$, and Figure~\ref{fig:e_d_01_05} shows the relations for $(\theta_{\rm c},\,I)=(0.1\pi,0.5\pi),\,(0.5\pi,0.5\pi)$.  
In the case of $e=0$, the duty cycle does not depend on $\omega$ because there is no specific direction for the circular orbit.
However, in the case of $e\neq0$, the duty cycle varies depending on the argument of periapsis.
In the case of $\left|\frac{3\pi}{2}-\omega\right|<\tilde{\theta_{\rm c}}$, the duty cycle grows with $e$ and approaches 100\% as $e$ approaches unity.
In the case of $\left|\frac{3\pi}{2}-\omega\right|>\tilde{\theta_{\rm c}}$, the duty cycle grows or decreases as $e$ grows, and finally approaches 0\% as $e$ approaches unity.
This is because the FRB pulsar spends more time around the apoapsis as $e$ grows, and the FRBs radiated at the apoapsis are observable in the case of $\left|\frac{3\pi}{2}-\omega\right|<\tilde{\theta_{\rm c}}$, and they are not observable in the case of $\left|\frac{3\pi}{2}-\omega\right|>\tilde{\theta_{\rm c}}$.
As these figures show, any duty cycle can be realized if the binary has the non-zero eccentricity and the viewing angles are free parameters.

\bibliographystyle{aasjournal}
\bibliography{cite.bib}
\end{document}